\documentclass[10pt, journal, letterpaper]{IEEEtran}

\usepackage[utf8]{inputenc}
\usepackage{epsfig}
\usepackage{amsmath}
\usepackage{graphicx}
\usepackage{amssymb}
\usepackage{fixmath}
\usepackage[belowskip=-5pt, font=small]{caption}
\usepackage{algorithmic}
\usepackage{algorithm}
\usepackage{multirow}
\usepackage[nolist,nohyperlinks]{acronym}
\usepackage{hyperref}
\usepackage[numbers,sort&compress]{natbib}
\usepackage{comment}
\usepackage[font=small]{subcaption}
\usepackage{xspace}
\usepackage{makecell}

\usepackage{dblfloatfix}

\usepackage{xcolor}

\makeatletter
\@namedef{ver@lineno.sty}{9999/12/31}
\@namedef{opt@lineno.sty}{}
\makeatother
\usepackage[newfloat, frozencache, cachedir=_minted-main]{minted}	

\usepackage{balance}

\acrodef{api}[API]{Application Programming Interface}
\acrodef{apn}[APN]{Access Point Name}
\acrodef{cellos}[CellOS]{Cellular Operating System}
\acrodef{cn}[CN]{Core Network}
\acrodef{cots}[COTS]{Commercial Off-the-Shelf}
\acrodef{csi}[CSI]{Channel State Information}
\acrodef{cqi}[CQI]{Channel Quality Indicator}
\acrodef{cu}[CU]{Central Unit}
\acrodef{du}[DU]{Distributed Unit}
\acrodef{epc}[EPC]{Evolved Packet Core}
\acrodef{gnbcu}[gNB-CU]{gNB Central Unit}
\acrodef{gnbdu}[gNB-DU]{gNB Distributed Unit}
\acrodef{ric}[RIC]{\acs{ran} Intelligent Controller}
\acrodef{gnbru}[gNB-RU]{gNB Radio Unit}
\acrodef{hss}[HSS]{Home Subscriber Server}
\acrodef{imei}[IMEI]{International Mobile Equipment Identity}
\acrodef{imsi}[IMSI]{International Mobile Subscriber Identity}
\acrodef{ip}[IP]{Infrastructure Provider}
\acrodef{lte}[LTE]{Long Term Evolution}
\acrodef{mac}[MAC]{Medium Access Control}
\acrodef{mme}[MME]{Mobility Management Entity}
\acrodef{mvno}[MVNO]{Mobile Virtual Network Operator}
\acrodef{oai}[OAI]{OpenAirInterface}
\acrodef{ofdm}[OFDM]{Orthogonal Frequency-division Multiplexing}
\acrodef{onap}[ONAP]{Open Network Automation Platform}
\acrodef{pdcp}[PDCP]{Packet Data Convergence Protocol}
\acrodef{pgw}[PGW]{Packet Data Network Gateway}
\acrodef{phr}[PHR]{Power Headroom}
\acrodef{phy}[PHY]{Physical}
\acrodef{prb}[PRB]{Physical Resource Block}
\acrodef{pucch}[PUCCH]{Physical Uplink Control Channel}
\acrodef{pusch}[PUSCH]{Physical Uplink Shared Channel}
\acrodef{qoe}[QoE]{Quality of Experience}
\acrodef{qos}[QoS]{Quality of Service}
\acrodef{pawr}[PAWR]{Platforms for Advanced Wireless Research}
\acrodef{pl}[PL]{partial linearization}
\acrodef{powder}[POWDER]{Platform for Open Wireless Data-driven Experimental Research}
\acrodef{ran}[RAN]{Radio Access Network}
\acrodef{renew}[RENEW]{Reconfigurable Eco-system for Next-generation End-to-end Wireless}
\acrodef{rlc}[RLC]{Radio Link Control}
\acrodef{rps}[RPS]{Reconfigurable Protocol Stack}
\acrodef{ree}[REE]{Reconfigurable Edge Element}
\acrodef{rnti}[RNTI]{Radio Network Temporary Identifier}
\acrodef{rrc}[RRC]{Radio Resource Control}
\acrodef{rs}[RS]{Reference Signal}
\acrodef{sdap}[SDAP]{Service Data Adaptation Protocol}
\acrodef{sdn}[SDN]{Software-Defined Networking}
\acrodef{sdr}[SDR]{Software-Defined Radio}
\acrodef{sgw}[SGW]{Serving Gateway}
\acrodef{sinr}[SINR]{Signal-to-Interference-plus-Noise Ratio}
\acrodef{sla}[SLA]{Service-Level Agreement}
\acrodef{snr}[SNR]{Signal to Noise Ratio}
\acrodef{tbs}[TBS]{Transport Block Size}
\acrodef{to}[TO]{Telco Operator}
\acrodef{ue}[UE]{User Equipment}
\acrodef{usrp}[USRP]{Universal Software Radio Peripheral}
\acrodef{zsm}[ZSM]{Zero-touch Network and Service Management}

\newcommand{\enb}{eNB\xspace}
\newcommand{\enbs}{\enb{}s\xspace}

\newcommand{\fig}[1]{Figure~\ref{fig:#1}}
\newcommand{\mathenb}[1]{$\mathrm{\enb{}_{#1}}$}

\renewcommand{\sec}[1]{Section~\ref{sec:#1}}
\newcommand{\ue}{\ac{ue}\xspace}
\newcommand{\ues}{\acp{ue}\xspace}
\newcommand{\cellos}{\acs{cellos}\xspace}

\newcommand{\telco}{\ac{to}\xspace}

\newcommand{\slice}[1]{Slice~#1}
\newcommand{\gnb}{gNB\xspace}
\newcommand{\powderrenew}{\acs{powder}-\acs{renew}\xspace}

\renewcommand{\bf}[1]{\mathbf{#1}}
\newcommand{\bs}[1]{\boldsymbol{#1}}

\graphicspath{{./figure/}}

\begin{document}

\title{CellOS: Zero-touch Softwarized Open\\Cellular Networks}

\author{Leonardo Bonati\IEEEauthorrefmark{1},
Salvatore D'Oro\IEEEauthorrefmark{1},
Lorenzo Bertizzolo\IEEEauthorrefmark{1},
\\Emrecan Demirors\IEEEauthorrefmark{1},
Zhangyu Guan\IEEEauthorrefmark{2},
Stefano Basagni\IEEEauthorrefmark{1},
Tommaso Melodia\IEEEauthorrefmark{1}\\
\vspace{4pt}
\IEEEauthorrefmark{1}Institute for the Wireless Internet of Things,\\Northeastern University, Boston, MA 02115, USA\\
\vspace{2pt}
\IEEEauthorrefmark{2}Department of Electrical Engineering,\\The State University of New York (SUNY) at Buffalo, Buffalo, NY 14260, USA\\
\vspace{2pt}
Email: \{l.bonati, s.doro, bertizzolo.l, e.demirors, s.basagni, t.melodia\}@northeastern.edu, guan@buffalo.edu
\thanks{This work was supported in part by the US National Science Foundation under Grant CNS-1618727 and in part by the US Office of Naval Research under Grants N00014-19-1-2409 and N00014-20-1-2132.}
}

\maketitle

\begin{abstract}
Current cellular networks rely on closed and inflexible infrastructure tightly controlled by a handful of vendors.
Their configuration requires vendor support and lengthy manual operations, which prevent Telco Operators (TOs) from unlocking the full network potential and from performing fine grained performance optimization, especially on a per-user basis.
To address these key issues, this paper introduces CellOS, a fully automated optimization and management framework for cellular networks that requires negligible intervention (``zero-touch'').
CellOS leverages softwarization and automatic optimization principles to bridge Software-Defined Networking (SDN) and cross-layer optimization.
Unlike state-of-the-art SDN-inspired solutions for cellular networking, CellOS: 
(i)~Hides low-level network details through a general \textit{virtual network abstraction};
(ii)~allows TOs to \textit{define high-level control objectives} to dictate the desired network behavior without requiring knowledge of optimization techniques, and 
(iii)~automatically generates and executes distributed control programs for simultaneous optimization of heterogeneous control objectives on multiple network slices.
CellOS has been implemented and evaluated on an indoor testbed with two different LTE-compliant implementations: OpenAirInterface and srsLTE.
We further demonstrated CellOS capabilities on the long-range outdoor POWDER-RENEW PAWR 5G platform.
Results from scenarios with multiple base stations and users show that CellOS is platform-independent and self-adapts to diverse network deployments. 
Our investigation shows that CellOS outperforms existing solutions on key metrics, including throughput (up to~86\% improvement), energy efficiency (up to~84\%) and fairness (up to~29\%).
\end{abstract}


\begin{picture}(0,0)(20,-540)
\put(0,0){
\put(0,20){\footnotesize This article has been published on Computer Networks. Please cite it as L. Bonati, S. D'Oro, L. Bertizzolo, E. Demirors, Z. Guan, S. Basagni, and T. Melodia,}
\put(0,10){\footnotesize ``CellOS: Zero-touch Softwarized Open Cellular Networks,'' Computer Networks, vol. 180, October 2020, doi: \href{https://doi.org/10.1016/j.comnet.2020.107380}{10.1016/j.comnet.2020.107380}.}
\put(0,0){\tiny \copyright 2020. This manuscript version is made available under the CC-BY-NC-ND 4.0 license \url{http://creativecommons.org/licenses/by-nc-nd/4.0/}}
\put(0,-5){\scriptsize}}
\end{picture}

\section{Introduction}

Current, state-of-the-art cellular networks rely on proprietary and inflexible hardware and software solutions produced and maintained by few vendors.
These closed architectures generally require manual configuration, preventing \acp{to} from being able to fully controlling resources such as spectrum, computing and transmission power to optimize network performance~\cite{radisys2019open, tip2019openran, kumar2014lte}.
Remedies to this fundamental limitation have been piecemeal, often based on offline solutions for frequency assignment and network planning~\cite{gonzalez2013optimization, siomina2012analysis}. 
Optimizing time-sensitive 
network functionalities also rests on heuristics
often engraved in the hardware fabric~\cite{korowajczuk2011lte, margolies2016exploiting}.
As of today, autonomous optimization of network parameters and swift and flexible control of real-time requirements of lower layer protocols are a territory that is largely uncharted.

Through \ac{sdn}, \acp{to} are breaking the imposed vendor lock-in by leaving the static and monolithic \ac{ran} architecture in favor of using a dynamically programmable, i.e., \emph{softwarized}, \textit{open \ac{ran}} for rapid and innovative network deployments~\cite{radisys2019open, tip2019openran, att2019recap, oran2020tip, oran2020grow}.
Although the benefits of such an open and multi-vendor approach have been showcased widely~\cite{oran2019plugfest},
how to fully embed softwarization in the future~5G infrastructure remains unsettled, as the highly dynamic and distributed nature of cellular networks is not amenable to be 
addressed by the centralized \ac{sdn} approach.
This issue is further exacerbated by the increasing densification of cellular deployments and users, which makes non-automated control ineffective, if feasible at all.
This is witnessed by recent works on cellular and wireless \ac{sdn} clearly lamenting that the swift dynamics of these networks generate an overwhelming amount of signaling traffic, hardly bearable by traditional softwarized controllers~\cite{thembelihle2017softwarization, foukas2016flexran, guan2018wnos, lynch2019automated}.
As a consequence, current hardware implementations and centralized softwarized approaches do not allow timely optimization of network behavior and the increasingly needed superior network performance~\cite{scutari2014decomposition, scutari2016parallel}.

\acp{to} are extremely sensitive to these issues. 
For example, the European Telecommunications Standards Institute (ETSI) formed the \acl{zsm} group to define fully-automated---\textit{zero-touch}---paradigms to provide flexibility to the highly decentralized technology of future wireless~\cite{etsi2019zsm}.
Similarly, the latest releases of the 3rd Generation Partnership Project~(3GPP)
include a functional split of 5G NR\footnote{Initially introduced as ``New Radio'' in~\cite{38913}, the term NR now generically refers to the 5G Radio Access Network, having lost its original meaning in the latest 3GPP specifications~\cite{38300}.} base stations (called gNBs) capabilities, so that network control decisions that involve large time scales are made at the \ac{gnbcu}, while lower layer and time-sensitive procedures are executed at the \acp{gnbdu} deployed closer to the users~\cite{3gpp.21.915}.
The Linux Foundation and the O-RAN Alliance are promoting and building the \ac{onap} and O-RAN, two automated orchestration frameworks to transition the rigid cellular infrastructure to an elastic and softwarized \textit{open \ac{ran}}~\cite{linux2018onap, oran2018oran}.
We observe that, although these approaches foresee network optimization as pivotal, they do not directly implement it.
As of now, this is left to the wits of the \telco and to the best of our knowledge there is no \textit{zero-touch} solution yet to perform it dynamically.

This paper contributes to the efforts toward \textit{automated softwarization} and \textit{self optimization} of future 5G networks by proposing \cellos, the first \textit{zero-touch} software framework for next-generation cellular networks.
Like an operating system interfacing hardware and software functions (whence the name), \cellos flexibly bridges \ac{sdn} with cross-layer distributed optimization techniques for the cellular architecture.
We push the \ac{sdn} paradigm beyond the traditional separation of control and data planes, in that we also decouple control from optimization, adding further and unprecedented flexibility.
Responding fully to ETSI requirements and industry interests, \cellos enables \textit{zero-touch control and optimization} of low-level network functionalities by providing \acp{to} with an efficient, automated, modular, and flexible network control platform.
Specifically, \cellos 
(i) allows \acp{to} to define centralized and high-level control objectives (e.g., ``maximize network throughput'') without requiring expertise in cross-layer optimization theory or knowledge of network specifics; 
(ii) provides a general \textit{virtual network abstraction} that shields the \telco from the complexity of a sophisticated framework by abstracting network infrastructure and parameters, including those known at \textit{run-time only} (e.g., user-to-base station associations and  channel information);
(iii) automatically converts high-level control directives into \textit{distributed cross-layer control programs} to be executed at each network edge element,
and 
(iv) enables \textit{zero-touch optimization of distinct control objectives on different network slices coexisting on the same infrastructure}~\cite{afolabi2018network}.
%

\begin{figure}[h]
\centering
\includegraphics[width=\columnwidth]{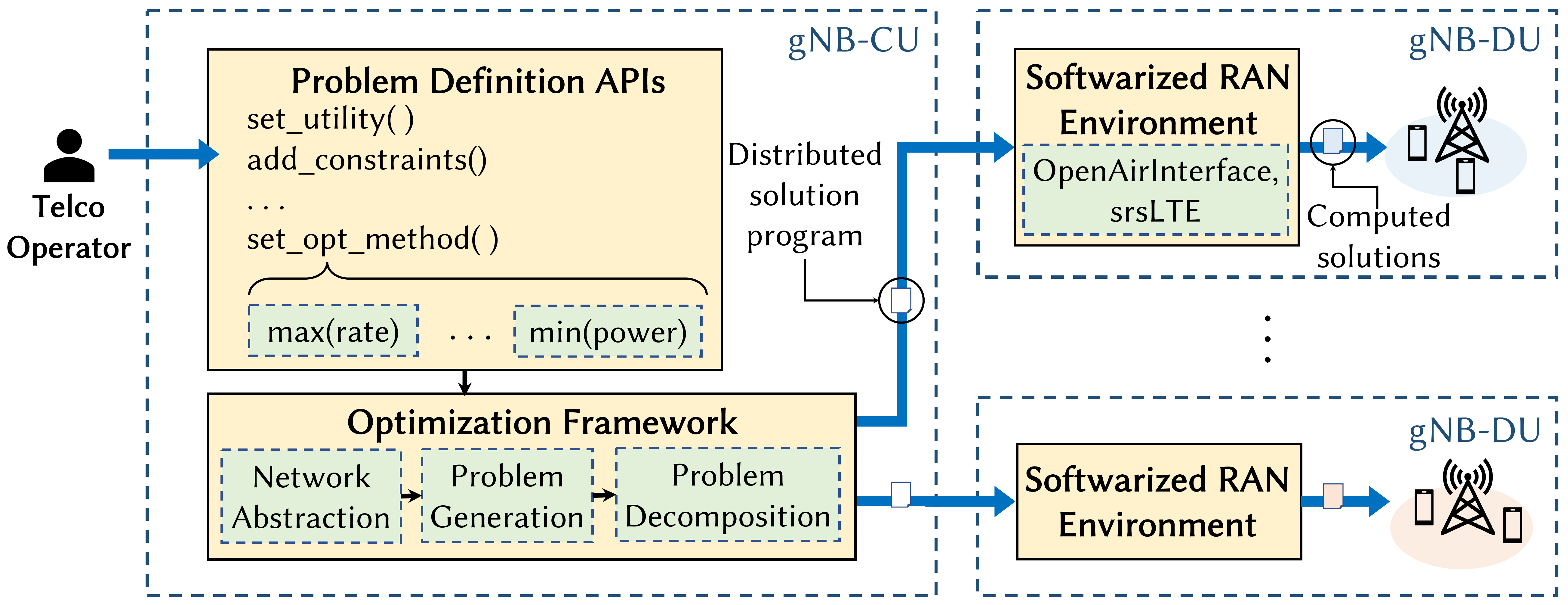}
\caption{\cellos at a glance as instantiated for the 3GPP architecture.}
\label{fig:architecture_high_level}
\end{figure}

\fig{architecture_high_level} illustrates the overall structure of \cellos, exemplified for the 3GPP network architecture. 

The upper-left side of the figure depicts the high level \acp{api} that the \acp{to} can use to define the network control objectives.
On the bottom we indicate the components of the framework for automatic generation of the optimization problems and their decomposition into control programs.
In a 3GPP scenario this unit corresponds to the \ac{gnbcu}, a logical node primarily concerned with control decisions at larger time-scales.
On the right, we describe the softwarized \ac{ran} that will execute the generated programs.
In the 3GPP context, this task would be carried out by the \ac{gnbdu}, a logical node that makes time-sensitive decisions involving the lower layers of the protocol stack, and that is interfaced with the \ac{gnbcu}.

We have prototyped \cellos on heterogeneous \ac{lte}-compliant testbeds.
We have chosen two different implementations of the \ac{lte} stack, namely, \ac{oai}~\cite{kaltenberger2020openairinterface} and srsLTE~\cite{gomez2016srslte}, to show that our framework is not tied to any specific \ac{ran} infrastructure.
Our experiments consider a variety of scenarios with multiple base stations and users to show that \cellos optimizes the network performance by swiftly adapting to varying network configurations and settings.
%
We also show the gains in performance that \cellos can bring to \ac{ran} implementations for cellular networks, such as \ac{oai} and srsLTE, as well as to \ac{mac}-layer scheduling algorithms commonly used in cellular networks, i.e., proportional fairness, greedy, and round-robin scheduling algorithms.
%
Results of the comparative performance evaluation of \cellos and prevailing baseline solutions show that using our framework remarkably improves key performance metrics, such as throughput (up to~86\%), energy efficiency (up to~84\%) and user fairness (up to~29\%).
We also show that \cellos is transparent to the use of network slicing technologies~\cite{doro2020slicing, doro2019slice, doro2020sledge}, enabling \acp{to} to simultaneously optimize different network functions on distinct network slices.
To the best our knowledge this is the first such demonstration, paving the way to the independent management of optimized network slices in 5G systems.
Finally, and for the first time, we provide evidence of the potentials of \textit{zero-touch optimization} in a \textit{softwarized \ac{ran}} ecosystem by testing \cellos on the long-range open-source \powderrenew \acs{pawr} 5G platform~\cite{pawr, powder}. 
Our results show that CellOS seamlessly interacts with the \ac{lte} protocol stack by optimizing resource allocation strategies, successfully increasing the average throughput by 23\%.

The remainder of the paper is organized as follows.
\sec{nutshell} presents \cellos in the 3GPP context, and a succinct overview of its main components.
Details of its architecture are provided in~\sec{framework}. An example of \cellos operations is given in \sec{examples}.
An \ac{lte}-compliant prototype of \cellos is illustrated in~\sec{prototype}.
\sec{results} reports the performance evaluation of \cellos on various testbeds, including a lab bench setup, the Arena testbed~\cite{bertizzolo2019arena}, and the \powderrenew \acs{pawr} 5G platform~\cite{pawr, powder}, using both the \ac{oai} and srsLTE \ac{ran} implementations with multiple base stations and users.
Work related to our research is surveyed in~\sec{related}.
Finally, \sec{conclusions} concludes the paper.

\section{CellOS in a 5G Flair}
\label{sec:nutshell}

This section provides a primer on 5G NR, and an overview of the main \cellos components and on how they can be integrated in the CU/DU functional split introduced by NR.

\subsection{A Brief Overview of 5G NR}

Compared to \ac{lte}, the 3GPP introduced a series of innovations in NR both in terms of layers of the protocol stack and functionalities, including the support for a wider range of carrier frequencies~\cite{mezzavilla2018end}.
The NR frame was endowed with a more flexible structure, which, although still being based on \ac{ofdm}, concerns a variable number of symbols per subframe and larger bandwidths with up to $400\:\mathrm{MHz}$ per carrier.
The 5G \ac{ran} can operate in two distinct configurations: \textit{Non-standalone}, i.e., paired with an \ac{lte} core network, and \textit{standalone}, i.e., connected to the new 5G Core.
Finally, NR base stations, called gNBs, can be deployed in a distributed manner across the network, dividing various parts of the NR protocol stack in different hardware components.

One of the main innovations that NR introduces is the split of the layers of the protocol stack of gNBs into distinct units.
These, namely \acf{gnbcu} and \acf{gnbdu}, can be deployed in separate locations across the cellular network~\cite{3gpp.21.915} (see \fig{architecture_high_level}).
Specifically, the \ac{gnbcu}, which can control multiple \acp{gnbdu}, involves the higher layers of the 3GPP protocol stack (i.e., \ac{pdcp}, \ac{sdap} and \ac{rrc}) and makes decisions at larger time scales.
The \ac{gnbdu}, instead, is deployed closer to the edge of the network and executes time-sensitive procedures, which involve the \ac{rlc}, \ac{mac}, and \ac{phy} layers of the protocol stack.
Moreover, the \ac{phy} layer of the \ac{gnbdu} can be additionally be broken down in a standalone \ac{gnbru}, which performs functions such as power amplification and signal transmission/reception~\cite{bonati2020open}.

While proposed by the 3GPP in~\cite{3gpp.38.816}, this separation has received significant attention due to O-RAN~\cite{oran2018oran}, which defined a series of interfaces between the aforementioned \gnb elements and a \ac{ric}, deployed at the edge of the network.
The \ac{ric} executes different functions of O-RAN, such as radio resource management, higher layers procedures and policy optimization, and control of \ac{ran} elements and resources.
Moreover, the \ac{ric} includes an application layer, which can host third-party components, such as \cellos, that regulate the behavior of the network.

\subsection{CellOS in a Nutshell}

A bird's-eye view of the \cellos architecture is shown in \fig{architecture_high_level}.
In line with the 3GPP \textit{functional split}~\cite{3gpp.21.915},  \cellos
is partitioned in \ac{gnbcu} and \ac{gnbdu} modular units to decouple the definition of network control procedures (at the \ac{gnbcu}) from their execution (at the \ac{gnbdu}).
\cellos main components are the interface to the \telco (providing the \emph{Problem Definition \acsp{api}}) and the automatic \emph{Optimization Framework} at the \ac{gnbcu}, and the \emph{Softwarized \ac{ran} Environment} at the \ac{gnbdu}.

By means of a rich variety of \acp{api}, the \telco sets the network control objective through high level, highly descriptive directives (e.g., ``maximize throughput''),
providing few key parameters (e.g., the number of base stations).
That is all the TO needs to specify, as \cellos abstracts the underlying network structure, hiding lower-level details to the \telco and mapping network elements such as base stations and \acp{ue} into virtual ones (\textit{Network Abstraction} block of our Optimization Framework).
As soon as the desired control objective is specified, \cellos converts it into a set of mathematical expressions that are used to define a centralized optimization problem, namely, the analytical representation of the optimization objective and of its constraints (\textit{Problem Generation} block in \fig{architecture_high_level}).
The generated problem is then \textit{automatically} decomposed into a set of distributed sub-problems, one for each of the edge elements (e.g., base stations).
This is done by the \textit{decomposition engine}, a core component of the \textit{Problem Decomposition} block.
Based on rigorous mathematical techniques, the centralized problem is partitioned both horizontally (decoupling variables belonging to different elements) and vertically (decoupling variables from different layers of each element's protocol stack).
The obtained sub-problems are then automatically converted into executable programs that are individually dispatched to each element (\textit{distributed solution programs}, in the \textit{Softwarized RAN Environment}).
%
%
Finally, each base station updates the distributed solution program with the real-time network parameters gathered from the \ac{ran} software stacks,
and runs it through its local solver.
It is worth mentioning that \cellos is independent of any specific \ac{ran} and can be interfaced with any other current or future 5G softwarized cellular stack. Finally, since \cellos edge elements have access to network real-time information by interfacing with the \ac{ran} software stacks (e.g., \ac{oai}, srsLTE), they update the received distributed control programs, adapting to the network time-varying dynamics, such as user arrival/departure, and mobility.
%

\section{CellOS Architecture}
\label{sec:framework}

%

\begin{figure*}[ht]
\setlength\belowcaptionskip{-10pt}
    \centering
    \includegraphics[width=\textwidth]{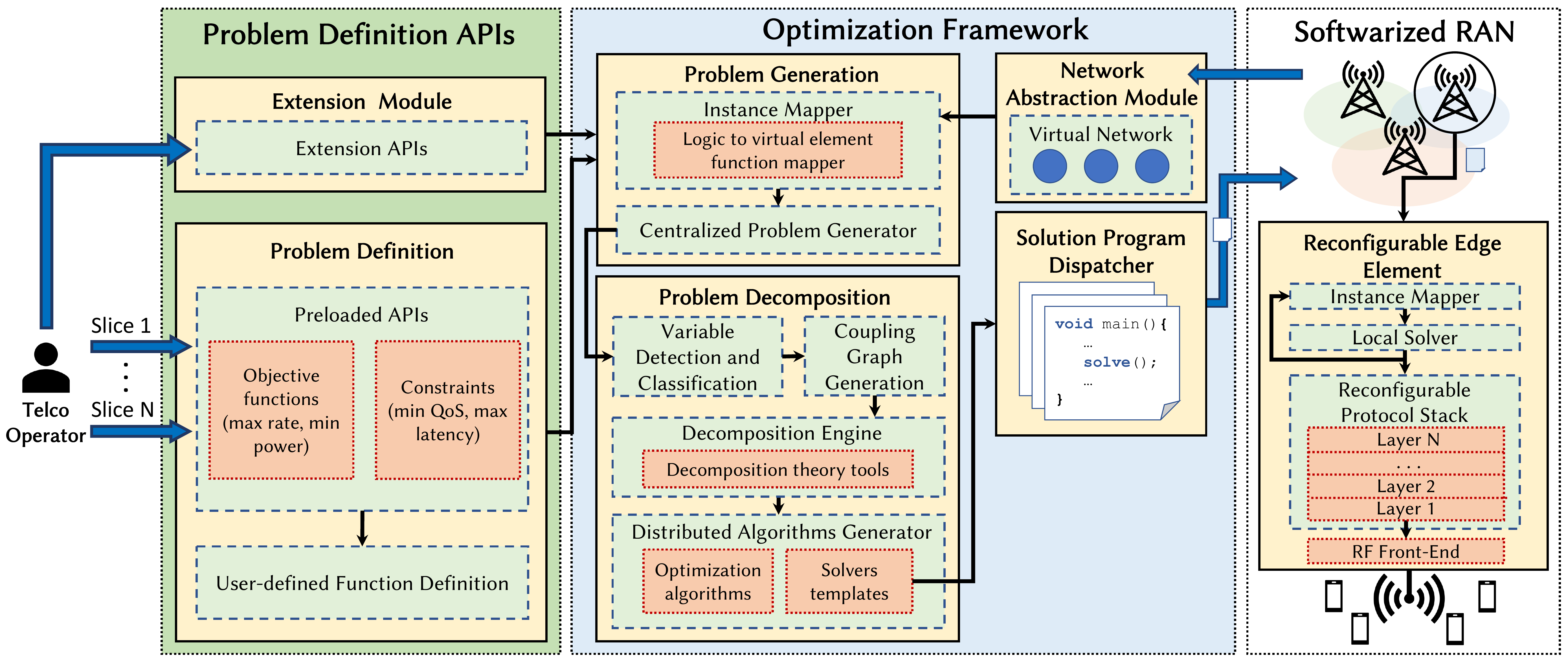}
    \caption{The \cellos architecture.}
    \label{fig:architecture}
\end{figure*}

In this section, we describe in details the components of the \cellos architecture, depicted in \fig{architecture}.

\subsection{Problem Definition APIs}
\label{sec:prob_definition}

\cellos defines a rich set of \acp{api}
to specify general high-level information about the desired network configuration and optimization.
These \acp{api} include functions to add base stations and for setting
per-user requirements (e.g., minimum rate guarantees).
%
The network control objective can be specified through a simple textual string, e.g., \textit{max(rate)} to maximize the network rate, \textit{min(power)} to minimize the overall power consumption.
%

\begin{listing}[ht]
\setlength\belowcaptionskip{-5pt}
\begin{minted}[style=vs,
baselinestretch=0.8,
frame=lines,
framesep=1mm,
fontsize=\scriptsize]{python}
1.  from cellos import Network, Engine

# Network instantiation
2.  nwk = Network(bs_num)
3.  slices = nwk.get_slices()

# Optimization problem and optional constraints definition
4.  nwk.set_utility('min(power)', slices[0])
5.  nwk.add_constraints({'user_min_rate':
                            [slices[0].get_users(), rate]})

# Optimization engine initialization
6.  eng = Engine()
7.  eng.set_opt_method('sub-gradient')
8.  nwk.initialize_engine(eng)
\end{minted}
\caption{\small{\cellos \acs{api} example.}}
\label{lst:api_sample}
\end{listing}
An example of \cellos \acp{api} and of the few lines of code needed to program a network objective are shown in Listing~\ref{lst:api_sample}.
In this example, the \telco instantiates a new network with a number \texttt{bs\_num} of base stations (line~2), and gets the network slices instantiated in the network (line~3).
An optimization problem aiming at minimizing the transmission power over a specific network slice (\texttt{slices[0]}) is then simply set in line~4, with constraints for guaranteeing a minimum rate defined in line~5.
It is worth noting that existing slices of the network, active subscribers, and associations of the two, are known a priori by the \telco, and stored, for instance, in the cellular core network.
We observe that very few lines of code are needed for the \telco to set the network goal, after which no further interaction is required.
This is because \cellos, dovetailed with the ETSI \textit{zero-touch} principles~\cite{etsi2019zsm}, hides all low-level network details (e.g., channel status, position of mobile users) from the \telco through the \textit{network abstraction module} (Section~\ref{sec:dispatcher}), and also automatically defines and distributively solves the optimization problem corresponding to the set control objective.

While specifying the objective function in textual form is enough for \cellos to properly work, experienced \acp{to} can define tailor-made custom and more advanced objective functions, optimization techniques, and solvers through an \textit{extension module}.
This provides additional \acp{api} for custom mathematical expressions and optimization constraints, and it also allows the \telco to select specific optimization techniques and solvers, as well as to achieve fine-grained control of network parameters and functionalities. These are then fed to the optimization framework in a way similar to the preloaded \acp{api}.
As of now, \cellos allows to specify functions expressed as linear combination of capacity, \ac{sinr}, power, and energy efficiency terms, which already enables \acp{to} to formulate a large number of wireless networking optimization problems~\cite{doro2015interference}.

\subsection{Optimization Framework}
\label{sec:prob_generation}

The heart of \cellos resides in its Optimization Framework, which:
(i) Converts the high-level centralized code into an optimization problem; (ii) decomposes it into sub-problems; (iii) creates and maintains an abstraction of the network, and (iv) dispatches the solution problems to the Softwarized \ac{ran}.

\subsubsection{Problem Generation}
\label{sec:subsec:problem_generation}

In order to transform high-level specifications into an optimization problem, \cellos first pairs high-level abstraction directives (control objective and constraints) with available network elements (e.g., base stations and users).
This is accomplished by the \emph{instance mapper module} that maps physical network elements to their virtual representation, and converts the control objective defined using high-level CellOS \acp{api} (\sec{prob_definition}) into machine-understandable code.
For example, \textit{max(sum(log(rate)))} is converted into $\max \sum_{u\in\mathcal{U}} \log(r_u)$, where $\mathcal{U}$ is the set of \acp{ue} and
$r_u$
their transmit rate.
%
%
The generated utility is kept as general as possible by using symbolic placeholders in lieu of parameters whose value will only be known at run-time (e.g., \ac{ue}-base station associations, channel coefficients, interfering signals, etc.).
In so doing, our Optimization Framework is \ue-agnostic.
It is the base stations that, at run-time, replace the symbolic placeholders with their current value.
Specifically, base stations interfaced with \cellos expose parameters and variables that can be tuned and optimized. Thus, placeholders of the generated problems always match physical network capabilities.

\subsubsection{Problem Decomposition}

This component of the Optimization Framework partitions the centralized problem into multiple sub-prob\-lems, one for each network element, to be
solved distributively at each base station.
%
%
In general, the centralized network control problem can be formalized as the following network utility maximization problem
%
\begin{align}
\underset{{\bf{x}}\in\mathcal{X}}{\text{maximize}} & \hspace{0.2cm} f(\bf{x}) \tag{CEN} \label{prob:num} \\
\text{subject to} & \hspace{0.2cm} g_i(\bf{x}) \leq h_i(\bf{x}), \quad \forall i\in\mathcal{I} \label{prob:constr}
\end{align}
where $\bf{x}$ represents the optimization variables (e.g., scheduling policies or transmission power levels), $\mathcal{X}$ is the strategy space (i.e., the set of all feasible strategy combinations), $f(\cdot)$ is the network-wide objective function (e.g., the overall capacity or the total energy efficiency of the network).
Inequality~\eqref{prob:constr} represents the set $\mathcal{I}$ of constraints (e.g., the transmission power must be bounded by some constant value; each \ac{prb} can be allocated to one \ac{ue} only, etc.). 
The biggest challenge in solving~\eqref{prob:num}
is that both objective function and constraints are, in general, coupled to different edge elements and to different layers of each element protocol stack.
Because of this tight coupling, generating distributed sub-problems that can be locally solved by each base station becomes challenging.

To address this challenge, \cellos first automatically identifies coupled variables and then applies rigorous decomposition 
to generate new sub-instances of~\eqref{prob:num} that are automatically assembled into uncoupled distributed programs to be executed at each base station.
This is accomplished performing the following (\fig{architecture}): \textit{variable detection and classification}, \textit{coupling graph generation}, decomposition (through the \textit{decomposition engine}), and \textit{distributed algorithms generation}.

\paragraph*{\textbf{Variable Detection and Classification}}%
\cellos starts by identifying the optimization variables of the network control problem.
This is done by parsing the generated objective function expression looking for symbolic placeholders introduced therein.
For instance, in~\eqref{prob:num} \cellos detects $\bf{x}$ to be the set of optimization variables of the problem.
Then, it determines which layer of the protocol stack houses which variable, e.g., power belongs to the \ac{phy} layer, scheduling to the \ac{mac} layer, and so on.
\cellos then identifies to which base station each variable belongs to.
As a result, each variable is assigned to a specific base station and to one of its protocol stack layers.

\begin{figure}[ht]
\setlength\belowcaptionskip{-5pt}
    \begin{center}
        \subcaptionbox{\label{fig:graph}}{\includegraphics[width=0.3\columnwidth]{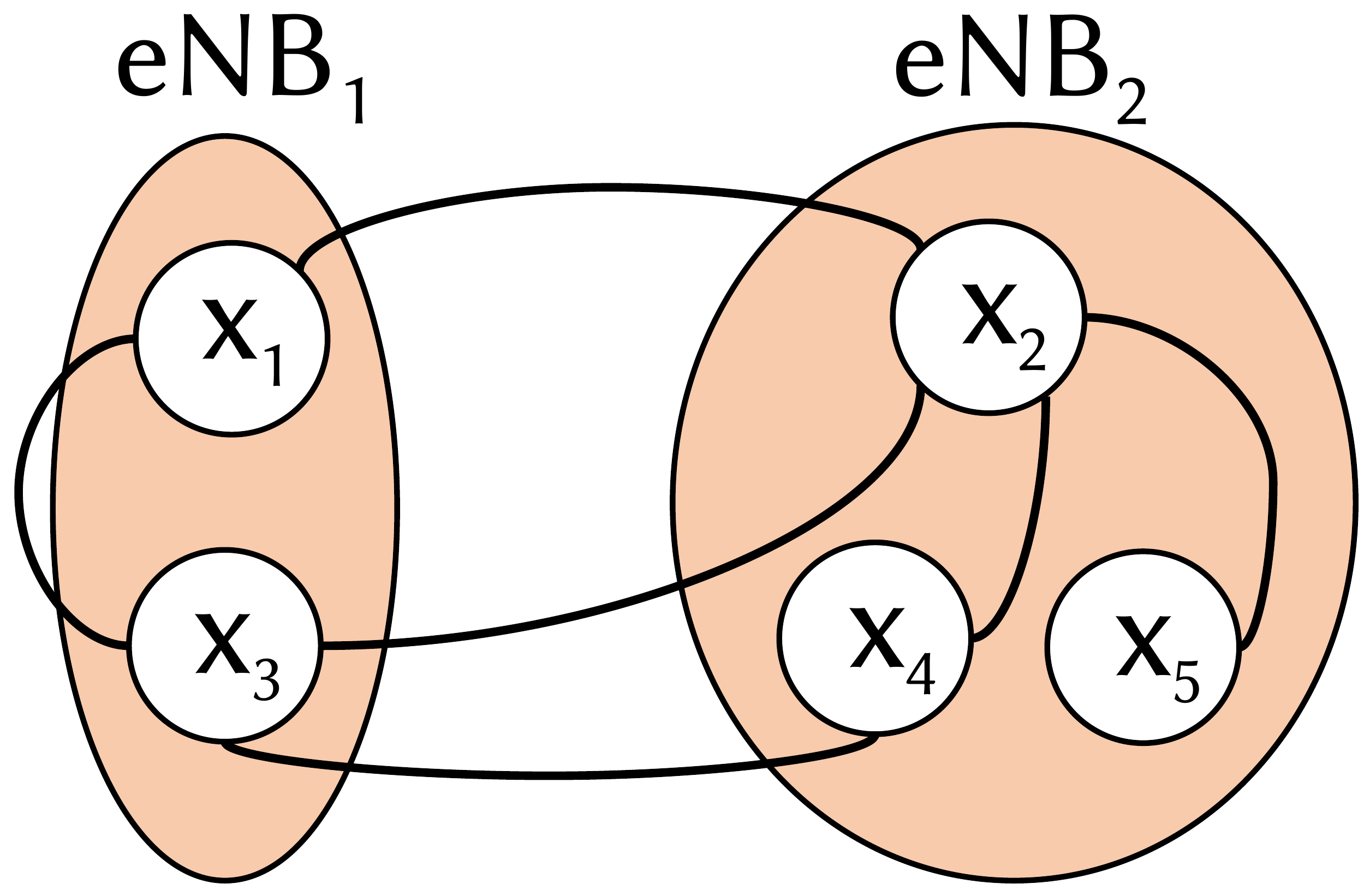}}\hspace{0.1\columnwidth}%
        \subcaptionbox{\label{fig:scenario_example}}{\includegraphics[width=0.29\columnwidth]{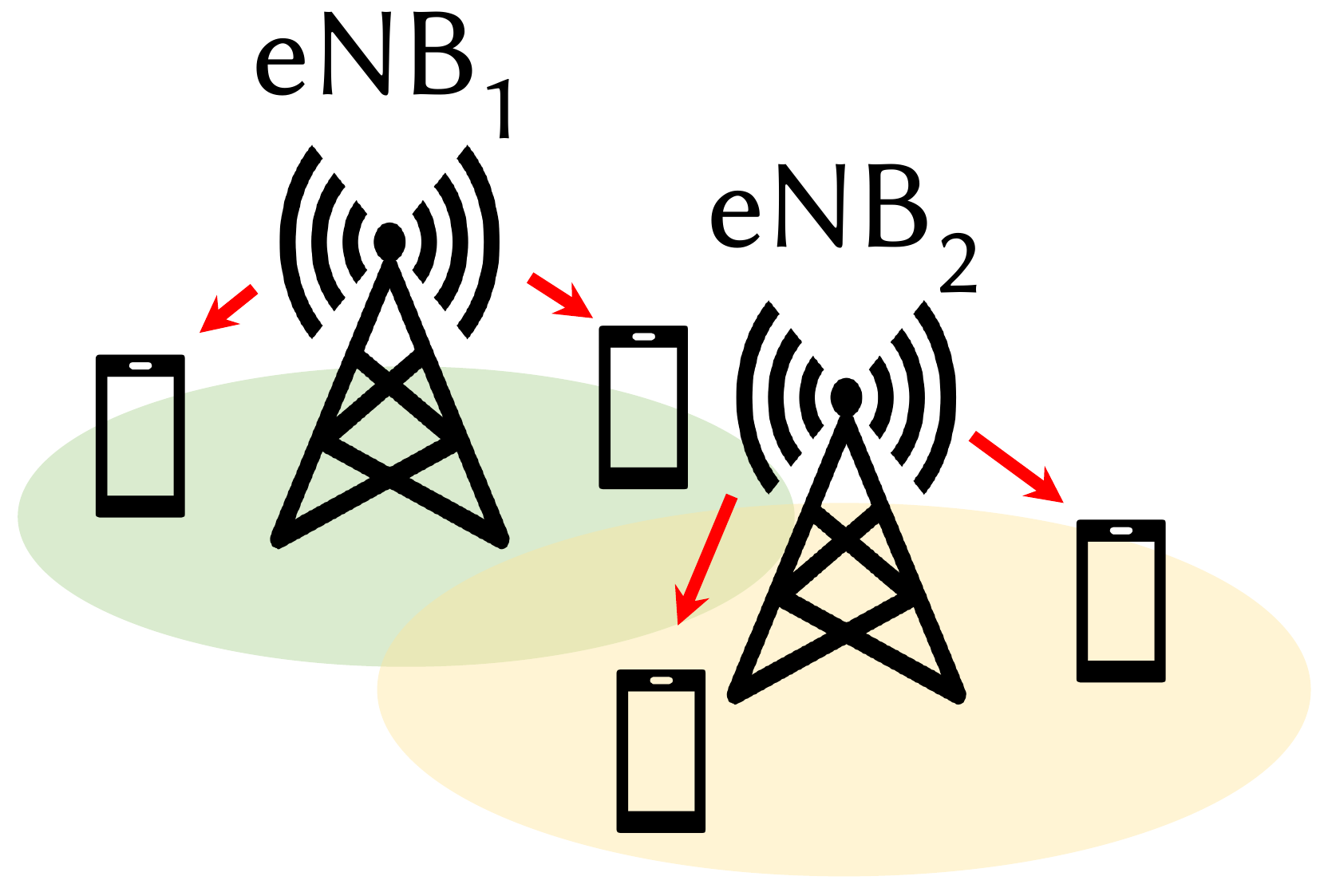}}
    \end{center}
    \caption{(a)~Coupling graph for $f(\mathbf{x}) = x_2(x_4+x_5)+x_3(x_4+x_1/x_2)$; (b)~Network scenario considered in \sec{examples}.}
\end{figure}

\paragraph*{\textbf{Coupling Graph Generation}}%
After detecting and classifying problem variables, \cellos organizes their coupling in a graph $G=(V,E)$, where $V$ is the set of variables of the network control problem, which are joined by an edge in $E$ only if they are coupled.
Similarly to what done in the previous step, coupling among variables is detected through a symbolic parser.
%
As an example, a coupling graph for $f(\mathbf{x}) = x_2(x_4+x_5)+x_3(x_4+\frac{x_1}{x_2})$ is shown in \fig{graph}.
Variables $\{x_i\}_{i=1,3}$ and $\{x_j\}_{j=2,4,5}$ belong to $\mathrm{eNB}_1$ and $\mathrm{eNB}_2$ (\fig{scenario_example}), respectively.

\noindent
\fig{graph} shows that coupling is not
limited to variables of a single \enb, but it might also involve those controlled by other \enbs.

\paragraph*{\textbf{Decomposition Engine}}
Variable detection/classification and coupling graphs are preliminary to automated problem decomposition, which we perform by using well-established techniques, including duality theory~\cite{bertsekas2015convex} and decomposition via \acl{pl}~\cite{scutari2014decomposition} (additional ones can be implemented through the \textit{extension module} of \fig{architecture}).
Decomposability is achieved introducing auxiliary variables (e.g., Lagrangian multipliers, penalization terms, and aggregate interference functions) that remove coupling across optimization variables and generate objective functions and constraints with separable terms in the sense of~\cite{bertsekas2015convex}.
Unfortunately, coupling in cellular networks involves heterogeneous network elements and different layers of the protocol stack, resulting in optimization problems whose utility or constraints are rarely separable.
For this reason, it is
classified into \textit{horizontal coupling} and \textit{vertical coupling}.
%
The former reflects dependencies among different network elements (e.g., among interfering base stations and their subscribers).
%
The latter, instead, concerns \textit{cross-layer} dependencies among different layers of the protocol stack of the same element (e.g., \ac{mac} policies affect transmission power and modulation strategies at the \ac{phy} layer).
Coupling makes centralized control of cellular networks extremely challenging as (i) the number of variables of the problem grows exponentially with the number of network elements, resulting in high computational and time complexity; (ii) the TO needs to be fully aware of the underlying network topology, the traffic demand, and the \ac{csi} for each individual \ue and base station, and (iii) centralized approaches require real-time information exchange between each network element and the centralized controller, imposing high signaling overhead and latency.
It is worth to point out that such network real-time information is not known at \cellos controller, but only at the edge elements.
Due to the fast changing network dynamics, though, the time required to signal local information to the controller, compute a centralized solution, and
adopt it at the edge elements might exceed the coherence time of the found solution.
\textit{Such solutions, may refer to an old network state and be obsolete, thus resulting in poor performance.}
This makes distributed solutions highly desirable, if not mandatory.
%
Even though distributed algorithms might not always guarantee globally optimal solutions, they usually manage to compute locally optimal ones with significantly lower computational complexity,
while ensuring run-time performance~\cite{scutari2014decomposition, scutari2016parallel}.
%

We point out that this work does not focus on
proposing new decomposition theories.
\textit{Our aim, instead, is to automatically generate distributed optimization programs based on a high-level objective, irrespective of the decomposition method used.}

\paragraph*{\textbf{Distributed Algorithms Generator.}}
The final step to achieve distributed control of the cellular network is to generate \emph{distributed solution programs} which can be executed and solved by each base station via standard
optimization solvers.
This
task is performed by the \textit{distributed algorithms generator} unit of \cellos Optimization Framework (\fig{architecture}).
As mentioned, the Optimization Framework is not cognizant of the value of parameters that are known at run-time only.
Accordingly, the \textit{distributed solution programs}
contain symbols in place of these parameters.
Each base station will then replace these symbols with their actual value at run-time, and associate optimization variables to the
served \ues.
The \textit{instance mapper} module has been designed to perform this task (\fig{architecture}).
This is one of the most important features of \cellos as it makes the solution program generation process (i) fully automated; (ii) independent of network configuration, and (iii) self-adapting to compute parameters at run-time based on current network conditions.

\subsubsection{Dispatcher and Abstraction Module}
\label{sec:dispatcher}

The
last two components of the Optimization Framework are the \textit{solution program dispatcher} and the \textit{network abstraction module}.
The \textit{dispatcher} utilizes sockets to transfer the generated distributed solution programs to each network base station, which will execute and solve them to achieve the desired network objective.
%

The \textit{network abstraction module} creates a high-level representation of the 
network infrastructure, hiding low-level, hardware/software details from the \telco.
This abstraction
allows the \textit{problem generation} (\sec{subsec:problem_generation}) to automatically convert
directives and constraints given through the \acp{api} of \sec{prob_definition} into mathematical expressions and utility functions.

\subsection{Softwarized RAN}
\label{sec:infrastructure}

The third main component of the \cellos architecture (\fig{architecture}) is in charge of running the distributed solution programs at each network element so as to reach the global network objective requested by the TO.
Once the dispatcher has delivered the programs, the \textit{instance mapper} component of the \ac{ree} replaces the symbolic placeholders in the program with their corresponding run-time values.
This component is capable of dynamically adapting solution programs to current network conditions, such as arrival/departure of \ues,
handovers, and \ac{csi}.
At the end of this mapping procedure each program is executed by the \textit{local solver} and a solution is computed.
As mentioned above, \cellos uses decoupling terms (e.g., Lagrangian multipliers) to allow individual base stations to coordinate with each other.
Relevant parameters are iteratively updated and exchanged among the coupled \acp{ree} through already available inter-base station interfaces (e.g., X2/Xn interfaces of cellular networks).

Since all the decisions are made locally at the base stations, at most $|\mathcal{U}| \, (|\mathcal{N}| \!+\! 1)$ variables need to be exchanged at each iteration, where $\mathcal{U}$ is the set of users, $\mathcal{N}$ are the available transmission channels, and $|\cdot|$ denotes the cardinality operator.
As we will demonstrate in \sec{scalability} through experimental results, this overhead is negligible if compared to that of centralized approaches, which need to gather local information at the central controller.
Because of this very limited signaling overhead, our framework effectively self-adapts to the network fast changing behavior.
Upon computing optimal solutions for each local network control problem (e.g., transmission and scheduling policies), these are used by each \ac{ree} through the \ac{rps}, which controls \ac{mac} and \ac{phy} layers, among others.

\section{CellOS in Action: An Example}
\label{sec:examples}

We consider the scenario depicted in
\fig{scenario_example}, where two interfering \enbs{} in the set $\mathcal{B}$ share two channels and serve two \ues each.
Here, $\mathcal{U}_b$ is the set of users $u$ served by \enb $b \in \mathcal{B}$.
We consider a downlink cross-layer optimization problem where each \enb has a
transmission power budget $P^{\rm{max}}$,
and that the \ues request a minimum capacity $C^{\rm{min}}$.
The optimization variables of this problem concern \ac{mac} and \ac{phy} layers, namely, user scheduling and transmission power allocation.
%
In this example, we assume that the \telco uses \cellos to maximize the network capacity.
The \telco first instantiates a network with two base stations (\texttt{nwk = Network(2)}).
Then the following network control objective is set on the slice controlled by the \telco: \texttt{nwk.set\_utility(`max(capacity)', slices[0])}.

On the other hand, \cellos needs to perform a more complex set of operations to reach the objective specified so succinctly by the TO.
Let $\bf{y} \!=\! ({\bf{y}}_1,{\bf{y}}_2)$ represent the network scheduling profile, where ${\bf{y}}_b \!=\! (y_{b,1,n},y_{b,2,n})_{n=1,2}$ is the scheduling profile for \enb $b\!\!\in\!\!\{1,2\}$.
Let $y_{b,u,n}$, instead, represent the scheduling variable such that $y_{b,u,n} \!=\! 1$ if user $u$ is scheduled for downlink transmission on channel $n \!\in \mathcal{N} \!=\! \{1,2\}$ and $y_{b,u,n} \!=\! 0$, otherwise.
Similarly, $\bf{p} \!=\! ({\bf{p}}_1,{\bf{p}}_2)$ represents the network power allocation profile, where ${\bf{p}}_b \!=\! (p_{b,1,n},p_{b,2,n})_{n=1,2}$ is the power allocation profile for \enb $b$, and $p_{b,u,n}$ represents the downlink transmission power from $b$ to user $u$ on channel $n$.
Let $C_{b,u,n}(\bf{y},\bf{p})$ be the capacity for \ue $u$ served by \enb $b$ on channel $n$, expressed as
%
\begin{equation}
\label{eq:capacity}
    C_{b,u,n}(\bf{y},\bf{p})  
    \!=\! B \log_2 \left(\!1 \!+\!\frac{g_{b,u,n} y_{b,u,n} p_{b,u,n}}{N\! +\! g_{b'\!\!,u,n}\!\! \sum\limits_{u'\!\in\mathcal{U}_{b'}} \!\!p_{b'\!\!,u'\!\!,n} y_{b'\!\!,u'\!\!,n}}\!\right)\!\!,
\end{equation}
\noindent
where $B$ is the employed bandwidth, $N$ is the background noise power, and $g_{b,u,n}$ is the channel gain coefficient computed by $u$ and sent to $b$, as part of standard cellular networks signaling procedures between user and base station (e.g., \ac{lte} \ac{pucch}).

The centralized network control problem can be expressed as the following Capacity Maximization Problem (CMP)
%
\begin{align}
\underset{{\bf{y}, \bf{p}}\in\mathcal{X}}{\text{maximize}} \hspace{0.2cm} & \sum_{b\in\mathcal{B}} \sum_{u\in\mathcal{U}_b} \sum_{n=1}^2 C_{b,u,n}(\bf{y},\bf{p}), & \tag{CMP} \label{prob:cap:ut} \\[-0.1cm]
\text{subject to} \hspace{0.2cm} & \sum_{n=1}^2 C_{b,u,n}(\bf{y},\bf{p}) \geq C^{\rm{min}}, \hspace{0.25cm} \forall b \in \mathcal{B}, u \in \mathcal{U}_b \label{prob:cap:c1} \\[-0.2cm]
& \hspace{-0.15cm} \sum_{u\in\mathcal{U}_b} \sum_{n=1}^2 p_{b,u,n} \leq P^{\rm{max}}, \hspace{0.45cm} \forall b \in \mathcal{B} \label{prob:cap:c2} \\[-0.1cm]
& \sum_{n=1}^2 y_{b,u,n} \leq 1, \hspace{1.55cm} \forall b \in \mathcal{B}, \forall u \in \mathcal{U}_b \label{prob:cap:c3}
\end{align}
\noindent
where~\eqref{prob:cap:c1} represents the users' minimum capacity constraint, \eqref{prob:cap:c2} enforces \enbs{}' power budget, and~\eqref{prob:cap:c3} guarantees that each \enb allocates each channel to a single \ue only.

The main challenges in decomposing \eqref{prob:cap:ut} are: (i) It is a Mixed Integer Non-Linear Programming problem, which is NP-hard in general~\cite{lee2011mixed}, and (ii)~both~\eqref{eq:capacity} and~\eqref{prob:cap:c1} are coupled among different \enbs{}.

\cellos recognizes $\bf{y}$ and $\bf{p}$ to be the problem optimization variables and associates them to the \ac{mac} and \ac{phy} layers, respectively.
Now, the \textit{problem decomposition} module understands which variables
belong to which \enb and creates a coupling graph similar to that in \fig{graph}.
This is, then, used to detect the
aggregate interference term in the capacity expression~\eqref{eq:capacity}. 
Accordingly, it defines the following auxiliary function
\begin{equation}
    h_{b,u,n}({\bf{y}}_{-b},{\bf{p}}_{-b}) = \sum_{b'\in\mathcal{B}\setminus\{b\}} g_{b',u,n} \sum_{u'\in\mathcal{U}_{b'}} p_{b',u'n} y_{b',u',n}
\end{equation}
\noindent
where ${\bf{y}}_{-b}=\bf{y}\!\setminus\!\{{\bf{y}}_b\}$ and ${\bf{p}}_{-b} = \bf{p}\!\setminus\!\{{\bf{p}}_b\}$ are the scheduling and power allocation variables of the \enbs belonging to \hbox{$\mathcal{B}\!\setminus\!\{b\}$}.
%
%
At this point, new auxiliary variables are introduced to rewrite~\eqref{prob:num} as
\begin{align}
\underset{{\bf{y},\bf{p}, \bf{i}}}{\text{maximize}} & \hspace{0.2cm} \sum_{b\in\mathcal{B}} \sum_{u\in\mathcal{U}_b} \sum_{n=1}^2 C_{b,u,n}({\bf{y}}_b,{\bf{p}}_b,{\bf{i}}_b), \tag{DCMP} \label{prob:cap:ut:dc} \\
\text{subject to} & \hspace{0.2cm} \sum_{n=1}^2 C_{b,u,n}({\bf{y}}_b,{\bf{p}}_b,{\bf{i}}_b) \geq C^{\rm{min}}, \hspace{0.2cm} \forall b \in\mathcal{B}, u \in\mathcal{U}_b \label{prob:cap:c1:dc} \\
& \hspace{0.2cm} i_{b,u,n} \geq h_{b,u,n}({\bf{y}}_{-b},{\bf{p}}_{-b}), \forall  b \in\mathcal{B}, u,n=1,2 \label{prob:cap:c2:dc} \\
& \hspace{0.2cm} \textit{Constraints} \hspace{0.2cm} \eqref{prob:cap:c2}, \eqref{prob:cap:c3} \nonumber 
\end{align}

\cellos can now use duality optimization tools to generate the following Lagrangian dual function

\vspace{-0.5cm}
\begin{align}
\label{eq:lagrangian}
    & L(\bs{\lambda}, \bs{\mu},\bf{i}, \bf{y}, \bf{p}) = \sum_{b \in \mathcal{B}} \sum_{u \in \mathcal{U}_b} \sum_{n=1}^2 C_{b,u,n}({\bf{y}}_b,{\bf{p}}_b,{\bf{i}}_b) \nonumber \\[-0.1cm]
    - & \sum_{b \in \mathcal{B}} \sum_{u \in \mathcal{U}_b} \lambda_{b,u} \left(C^{\rm{min}} - \sum_{n=1}^2 C_{b,u,n}({\bf{y}}_b,{\bf{p}}_b,{\bf{i}}_b)\right) \nonumber \\[-0.1cm]
    - & \sum_{b \in \mathcal{B}} \sum_{u \in \mathcal{U}_b} \sum_{n=1}^2 \mu_{b,u,n} \left(h_{b,u,n}({\bf{y}}_{-b},{\bf{p}}_{-b}) - i_{b,u,n}\right),
\end{align}
\noindent
where $\bs{\lambda}=(\lambda_{b,u,n})$ and $\bs{\mu}=(\mu_{b,u,n})$ are the non-negative Lagrangian multipliers used in constrained optimization~\cite{bertsekas2015convex}.

We observe that problems~\eqref{prob:cap:ut} and~\eqref{prob:cap:ut:dc}, and the Lagrangian dual function~\eqref{eq:lagrangian} all aim at solving the centralized control problem~\eqref{prob:num}.
However, the advantage of using~\eqref{eq:lagrangian} is that function $L(\bs{\lambda}, \bs{\mu},\bf{i}, \bf{y}, \bf{p})$ is written with separable variables, meaning that it can be split into $|\mathcal{B}|$ sub-problems locally solvable by each \enb.

Finally, \cellos dispatches the generated distributed solution programs to the \enbs{} that populate them with network run-time information
(e.g., users' channel coefficients), and compute optimized solutions through their \textit{local solver}.

It is worth noting that the procedures detailed in Sections~\ref{sec:prob_definition} and~\ref{sec:prob_generation} need to be executed only once per control problem specified by the \telco and that they take very little time to be performed, e.g., $0.03\:\mathrm{s}$ for the example of this section (more details on the scalability of \cellos automatic procedures will be given in \sec{scalability}).

\section{OAI-based CellOS Prototype}
\label{sec:prototype}

\begin{figure}[ht]
    \centering
    \includegraphics[width=\columnwidth]{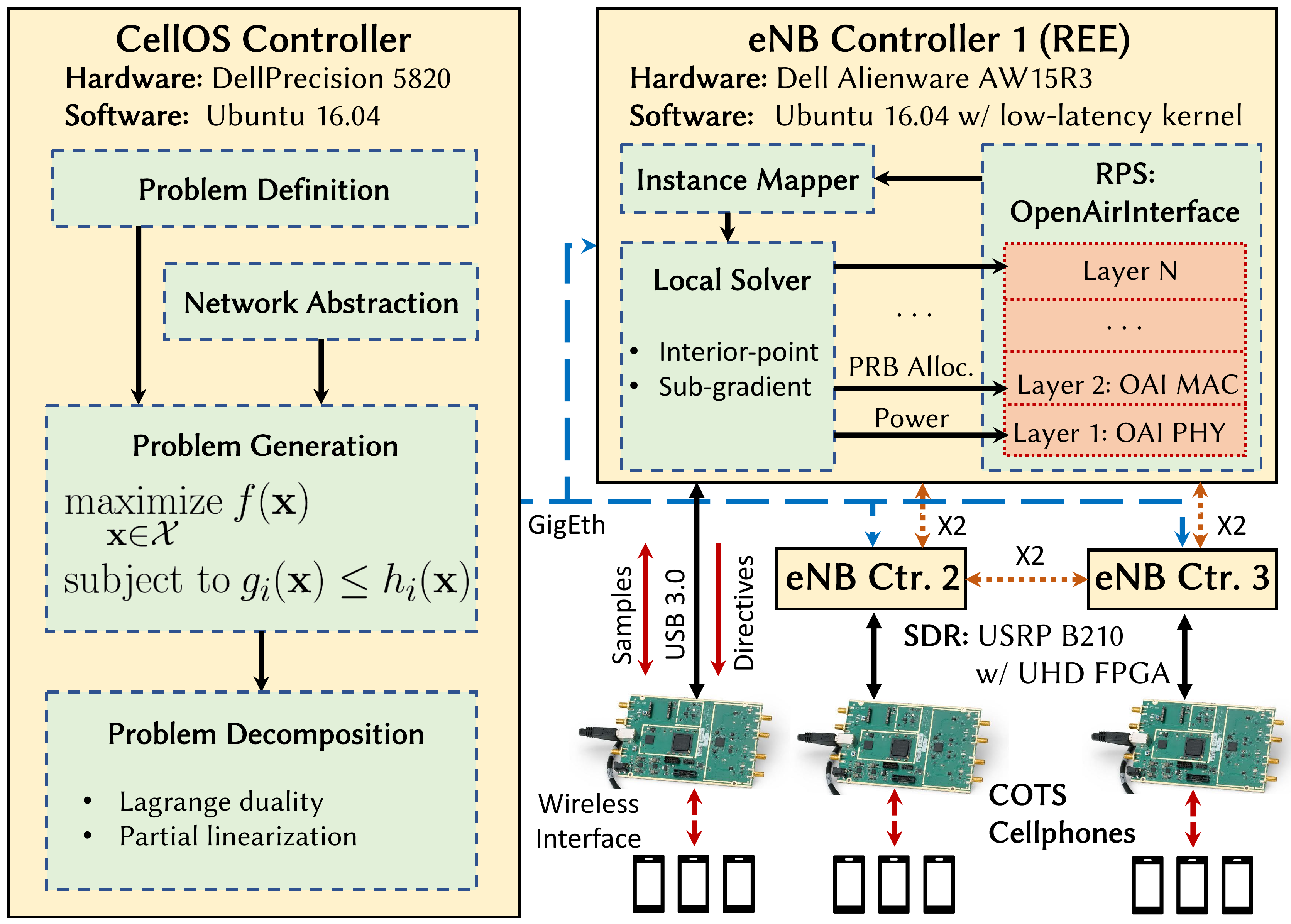}
    \caption{OAI-based \cellos prototype.}
    \label{fig:prototype}
\end{figure}

In this section, we discuss the prototypes of \cellos, which have been built based on the \ac{oai} and srsLTE open-source \ac{ran} implementations.
The \acs{oai}-based prototype is illustrated in \fig{prototype}.

The \textit{CellOS Controller}
performs the functionalities of the Problem Definition \acp{api} and of the Optimization Framework.
Particularly, it creates and maintains the network abstraction, generates the optimization problem based on the directives from the \telco, and
%
performs the problem decomposition.
In our experiments the decomposition process is obtained through
Lagrangian duality theory~\cite{bertsekas2015convex} and decomposition via \acl{pl}~\cite{scutari2014decomposition}.

Multiple \emph{\enb Controllers}, one for each base station, are connected to the \cellos Controller through a
Gigabit Ethernet connection.
These controllers use interior-point and sub-gradient algorithms \cite{bertsekas2015convex} to solve the received distributed programs,
and set the parameters to be used with the RF front-ends they are connected to.  
Each of these controllers drives an Ettus Research \ac{usrp} B210, which
serves \acp{ue} over \ac{lte} frequencies.
%
%
%
As \acp{ue} we used a set of heterogeneous \ac{cots} cellular phones (Samsung Galaxy S5, S6 and S7, and Apple iPhone 6s).
%

In this prototype, \cellos interfaces with the \ac{lte} protocol stack implementation offered by \textit{\acl{oai}}, i.e., an open-source software-based experimental platform for \ac{lte} implementations~\cite{kaltenberger2020openairinterface}.
%
\ac{oai} features \ac{lte} \ac{ran} applications along with \acl{epc} components.
As OAI does not directly allow per-user power control, or optimized PRB allocation---key essential requirements of many network control objectives---we have extended its functionalities by significantly modifying its core implementation.
%
Specifically, power control is obtained by amplitude-modulating the downlink data signal intended for a specific \ue.
%
%
\ac{prb} allocation, instead, is based on an optimized waterfilling-like fair scheduling algorithm~\cite{scutari2009mimo}, which has low-complexity, thus complying with \ac{lte} strict timing requirements.
Because of the \ac{prb} short time duration it is of utmost importance to compute the \ac{prb} allocation very quickly to guarantee compliance with \ac{lte} and promptly serve the \ues.
According to our scheme, \acp{prb} are allocated only to those UEs whose downlink transmission buffer is not empty.

A similar approach has been followed for the \textit{srsLTE} prototype, which leverages \acp{usrp} X310 in place of \acp{usrp} B210. This time, each \enb controller connects to the \ac{sdr} through a $10\:\mathrm{Gbit/s}$ PCI Express network card.
In this prototype, \cellos interfaces with the open-source cellular protocol stack offered by srsLTE, which, similarly to what done for \ac{oai}, has been extended to perform \ac{phy}-layer power control by adjusting the \acp{usrp} transmission power, and \ac{mac}-layer scheduling by optimally allocating \acp{prb} to \ues.

\section{Experimental Evaluation}
\label{sec:results}

The effectiveness of \cellos in automatically creating distributed optimization programs from high-level directives, and in managing the network infrastructure to reach different control objectives, is demonstrated via experimentation on various \ac{lte}-compliant testbeds.
We describe our testbed in \sec{testbed}, we introduce the investigated performance metrics in \sec{metrics}, and present our experimental results in \sec{evaluation}.

%

\subsection{Network Scenarios and Testbed Settings}
\label{sec:testbed}

To demonstrate its platform-independence, we test \cellos over different software and hardware platforms, using \ac{oai} and srsLTE, as well as heterogeneous software-defined radios and testbeds.

\begin{figure}[ht]
    \begin{center}
        \subcaptionbox{\label{fig:high_interference_scenario}High interference.}{\includegraphics[width=0.4\columnwidth]{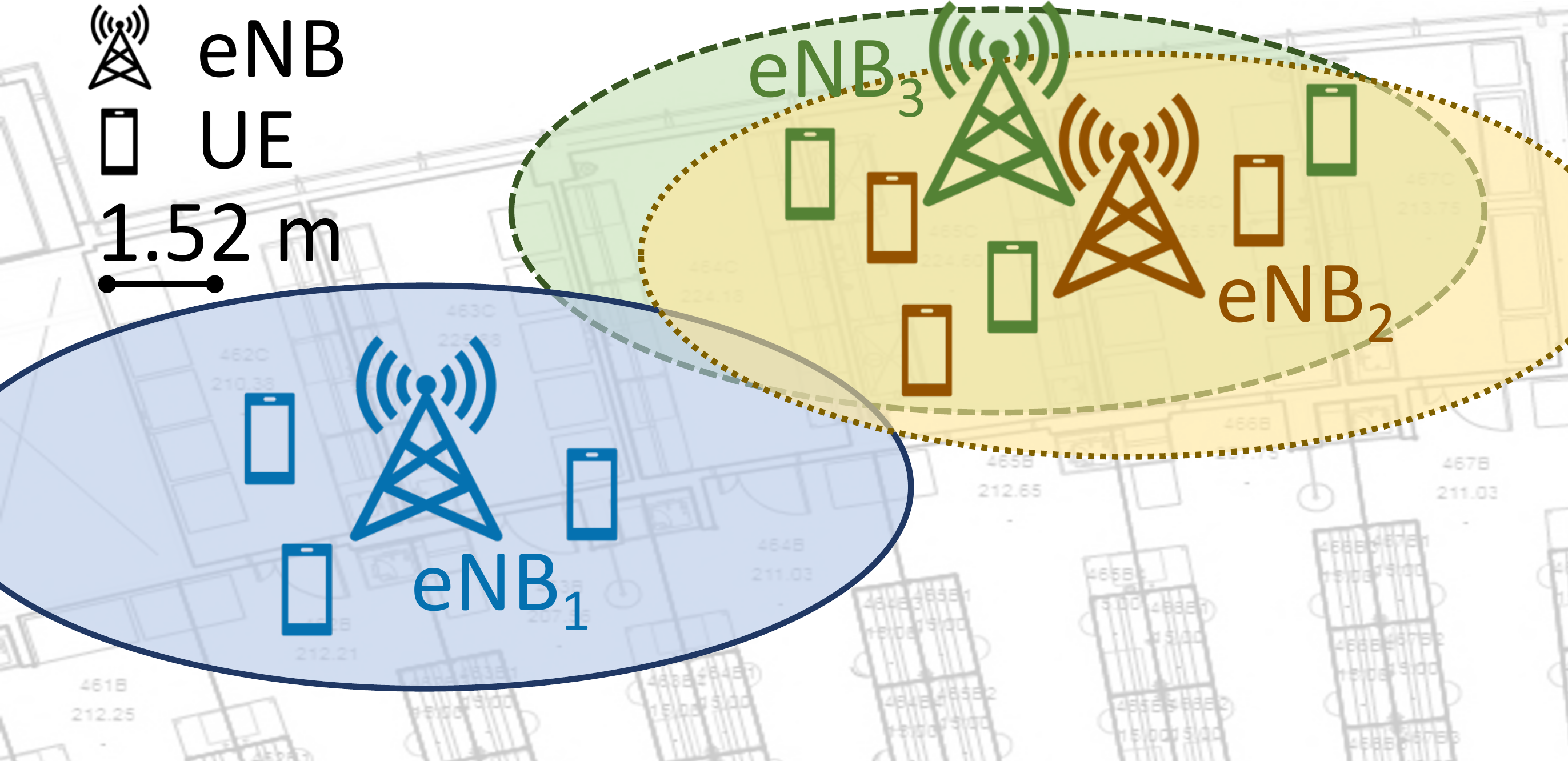}}\hspace{0.07\columnwidth}%
        \subcaptionbox{\label{fig:low_interference_scenario}Low interference.}{\includegraphics[width=0.4\columnwidth]{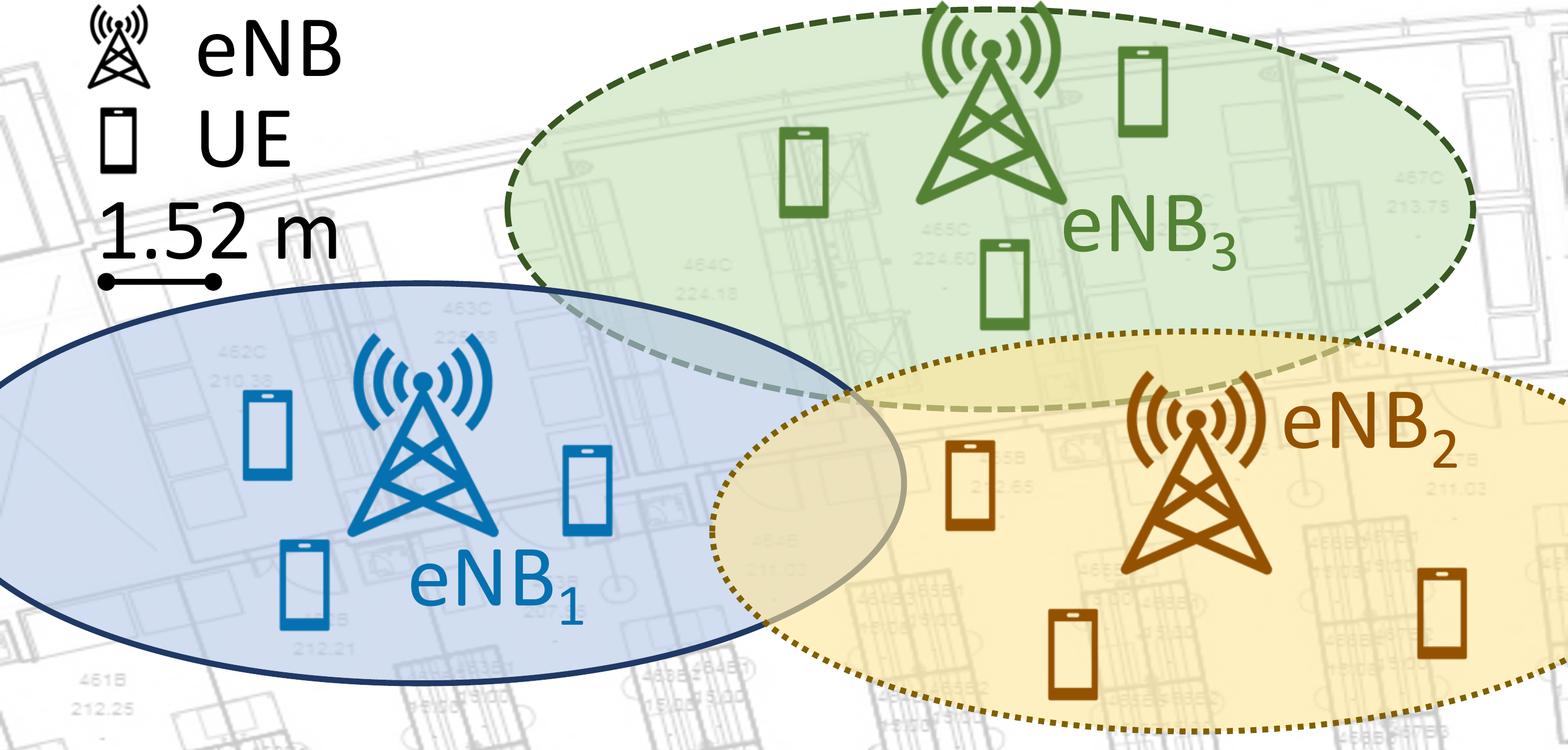}}
    \end{center}
    \caption{The CellOS lab bench testbed.}
    \label{fig:testbed}
\end{figure}

The \acs{oai}-based prototype
of Section~\ref{sec:prototype} has been used in a testbed composed of~$3$~\enbs{} and
up to~$9$~\acp{ue}. 
%
%
Each \enb uses a $10\:\mathrm{MHz}$ channel bandwidth corresponding to 50~\acp{prb}.
%
For this prototype we consider the two indoor scenarios depicted in \fig{testbed}:
(i) A high interference scenario, where 
two \enbs are in line-of-sight conditions and have largely overlapping coverage areas (\fig{high_interference_scenario}), and (ii) a low interference scenario where \enbs are in non-line-of-sight conditions and their coverage areas only partially overlap with each other  (\fig{low_interference_scenario}).

The high interference scenario represents those crowded environments (e.g., university campuses, concert halls or convention centers) where several femtocells are deployed in a crowded region to balance the traffic load of a macrocell farther away.
In this case, while the interference among macro- and femtocells is small, femtocells with overlapping coverage areas are subject to significant inter-cell interference.
In the low interference scenario, instead, \enbs{} are located far away from each other and, thus, are less subject to inter-cell interference and the subsequent performance degradation.

%
The srsLTE-based prototype is evaluated on a low-interference setup on the Arena testbed~\cite{bertizzolo2019arena}.
We instantiated 3~\ac{lte} \enbs on \acp{usrp} X310 whose antennas are connected to the ceiling of a $208.1\:\mathrm{m^2}$ office space.
A set of Dell EMC PowerEdge R340 servers are used to drive the \acp{usrp} through $10$~Gigabit Ethernet connections.
%
%
%
This set of experiments
shows that \cellos can \textit{simultaneously obtain different control objectives on multiple network slices.}
This represents the scenario in which multiple \acp{mvno} share the same edge elements, or that of a single \ac{to} wishing to set diverse control problems on each network slice.
%
To demonstrate the benefits of automatic optimization of the \textit{open \ac{ran}}, we finally instantiate \cellos on the long-range open-source 5G \powderrenew platform~\cite{powder}, which is the combination of the \acf{powder} and \acf{renew}, and part of the \acf{pawr}~\cite{pawr}.



We assess \cellos performance by letting \ues download a file stored on our local server for $60\:\mathrm{s}$.
%
\textit{It is worth mentioning that it only took \cellos $1.43\:\mathrm{s}$ and 8~lines of code (see Listing~\ref{lst:api_sample}) to automatically generate the evaluated distributed control programs} (more details on the scalability of these operations will be given in \sec{scalability})
%

\subsection{Performance Metrics} 
\label{sec:metrics}

\cellos has been evaluated against the following metrics.

\begin{itemize}
    \item \textit{Sum throughput of the network}, defined as  
    
\begin{equation}
\label{eq:throughput}
    S = \sum_{b \in \mathcal{B}} \sum_{u \in \mathcal{U}_b} S_{b, u}, \quad \forall b \in \mathcal{B}, u \in \mathcal{U}_{b}
\end{equation}

\noindent
where $\mathcal{B}$ and $\mathcal{U}_{b}$ are the sets of the \enbs{}~$b$ and of \acp{ue}~$u$ they are serving, and $S_{b, u}$ is the throughput offered to~$u \in \mathcal{U}_{b}$ by~$b$.
    
\item \textit{Normalized transmission power} of the base stations to the \ues. 
To analyze the impact of power control policies on the transmission power of \enbs{},  we show the transmission power of the base stations normalized to their maximum transmission power.
Let~$P^{\rm{max}}_b$ and~$P^{\rm{min}}_b$ be the maximum and minimum power levels of base station~$b$, the normalized transmission power is defined as
\begin{equation}
     P^N_{b,u} = \frac{P_{b,u} - P^{\rm{min}}_b}{P^{\rm{max}}_b - P^{\rm{min}}_b}, \quad \forall b \in \mathcal{B}, u \in \mathcal{U}_{b}
\end{equation}

\noindent
where $P_{b,u} \in \{ P^{\rm{min}}_b, P^{\rm{max}}_b \}$ is the power used by \enb $b \in \mathcal{B}$ to transmit to its user $u \in \mathcal{U}_{b}$.
    
\item \textit{Global energy efficiency}, defined as the amount of information per unit of energy the \enbs{} transmit to their subscribers:
\begin{equation}
\label{eq:energy_efficiency}
    EE = \frac{\sum_{b \in \mathcal{B}} \sum_{u \in \mathcal{U}_{b}} S_{b,u}}{\sum_{b \in \mathcal{B}} \sum_{u \in \mathcal{U}_{b}} P_{b,u}}, \quad \forall b \in \mathcal{B}, u \in \mathcal{U}_{b}
\end{equation}

\noindent
where $P_{b,u}$ is the power used by \enb~$b$ to transmit to its user~$u$.
    
\item \textit{System fairness}, measured through Jain's equation~\cite{jain1984quantitative}.
%
%
Given users $u \in \mathcal{U} = \bigcup_{b \in \mathcal{B}} \mathcal{U}_{b}$, Jain's fairness index $\mathcal{J}$ is defined as
\begin{equation}
\label{eq:fairness_index}
     \mathcal{J} = \frac{(\sum_{b \in \mathcal{B}} \sum_{u \in \mathcal{U}_{b}} S_{b,u})^2}{|\mathcal{U}| \sum_{b \in \mathcal{B}} \sum_{u \in \mathcal{U}_{b}} S_{b,u}^2}, \quad \forall b \in \mathcal{B}, u \in \mathcal{U}_{b}.
\end{equation}
\end{itemize}

\subsection{Experimental Results}
\label{sec:evaluation}


\begin{table}[ht]
\centering
\caption{Summary of experimental setup.}
\label{tab:experiment_setup}
\footnotesize
\setlength{\tabcolsep}{1pt}
\footnotesize
\renewcommand{\arraystretch}{1.6}
\begin{tabular}[]{|l|c|c|c|c|}
\hline
\textbf{Figure}    & \textbf{\makecell{Optimization\\Problem}}  & \textbf{Scenario}   & \textbf{\makecell{\acs{ran}\\Software}} & \textbf{Testbed} \\
\hline
Fig.~\ref{fig:maxRate}   & max(rate)    & \multirow{3}{*}{High Interference}  & \multirow{5}{*}{\acs{oai}~\cite{kaltenberger2020openairinterface}} & \multirow{5}{*}{\makecell{Lab\\Bench\\Setup}} \\
\cline{1-2}
Fig.~\ref{fig:high_interference_min_power_throughput}    & min(power)   &  &  & \\
\cline{1-2}
Fig.~\ref{fig:bar_plots} & max(sum(log(rate)))  &    &  & \\
\cline{1-3}
Fig.~\ref{fig:low_interference_sumlog_rate_throughput}   & max(sum(log(rate)))   & \multirow{2}{*}{Low Interference} &  & \\
\cline{1-2}
Fig.~\ref{fig:low_interference_scheduling}   & max(rate) &  &  & \\
\hline
Fig.~\ref{fig:slice_max_rate}    & max(rate) & \multirow{2}{*}{Slicing}   & \multirow{2}{*}{srsLTE~\cite{gomez2016srslte}} & \multirow{5}{*}{Arena~\cite{bertizzolo2019arena}} \\
\cline{1-2}
Fig.~\ref{fig:slice_min_power}   & min(power)    &   &  &  \\
\cline{1-4}
Fig.~\ref{fig:scalability}   &  \multirow{3}{*}{\makecell{max(rate),\\min(power),\\max(sum(log(rate)))} }  & \makecell{Controller Time} & \multirow{3}{*}{N/A} & \\
\cline{1-1}\cline{3-3}
Fig.~\ref{fig:local_solver}  &   & \makecell{Local Solver Time}  & & \\
\cline{1-1}\cline{3-3}
Fig.~\ref{fig:overhead}  &   & Signaling Overhead  & & \\
\hline
Fig.~\ref{fig:powder_thr}    & max(rate) & Long-range    & srsLTE~\cite{gomez2016srslte} & \makecell{POWDER-\\RENEW\\\cite{powder, pawr}} \\ 
\hline
\end{tabular}
\end{table}

\begin{figure*}[tb]
\begin{center}
\begin{minipage}{\textwidth}
\begin{minipage}{0.7\textwidth}
\setlength\abovecaptionskip{0cm}
    \begin{center}
        \subcaptionbox{\label{fig:high_interference_max_rate_9ues}OAI w/ and w/o \cellos.}{\includegraphics[height=4.1cm]{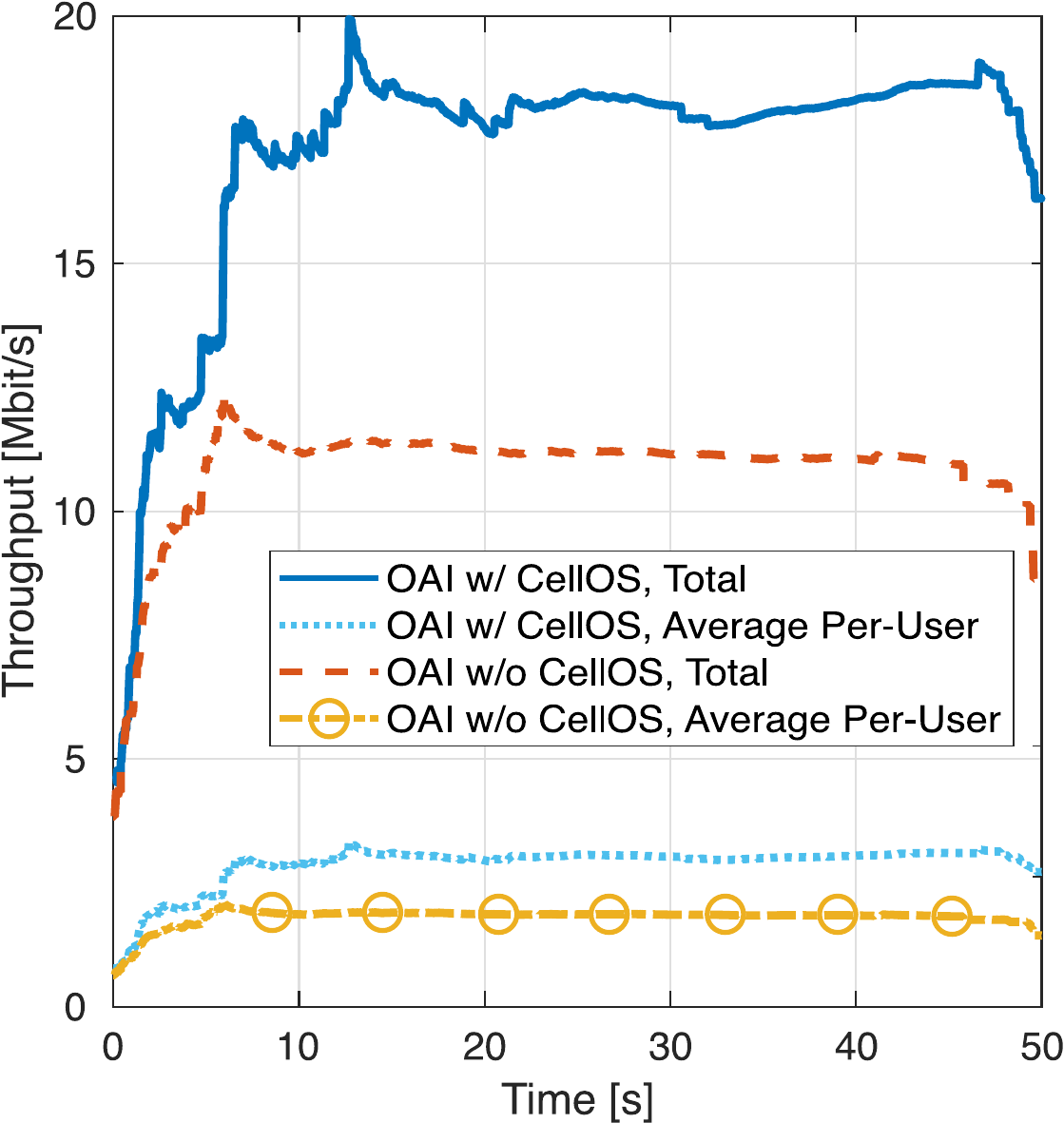}}\hspace{5pt}%
        \subcaptionbox{\label{fig:high_interference_max_rate_prb}\acsp{prb} at time $t_1$ and $t_2$.}{\includegraphics[height=4.1cm]{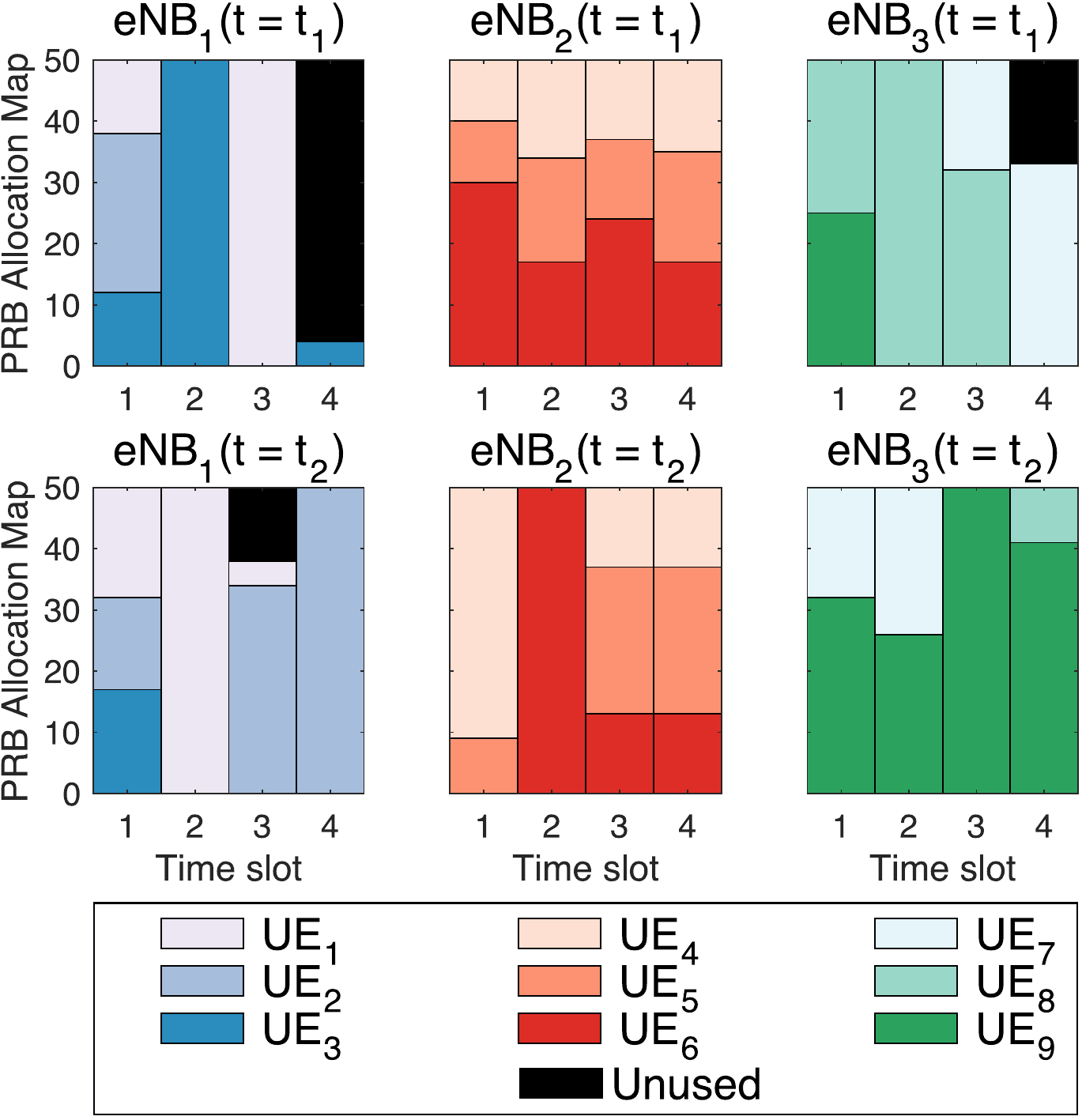}}
        \subcaptionbox{\label{fig:high_interference_max_rate_throughput}Throughput and power.} {\includegraphics[height=4.1cm]{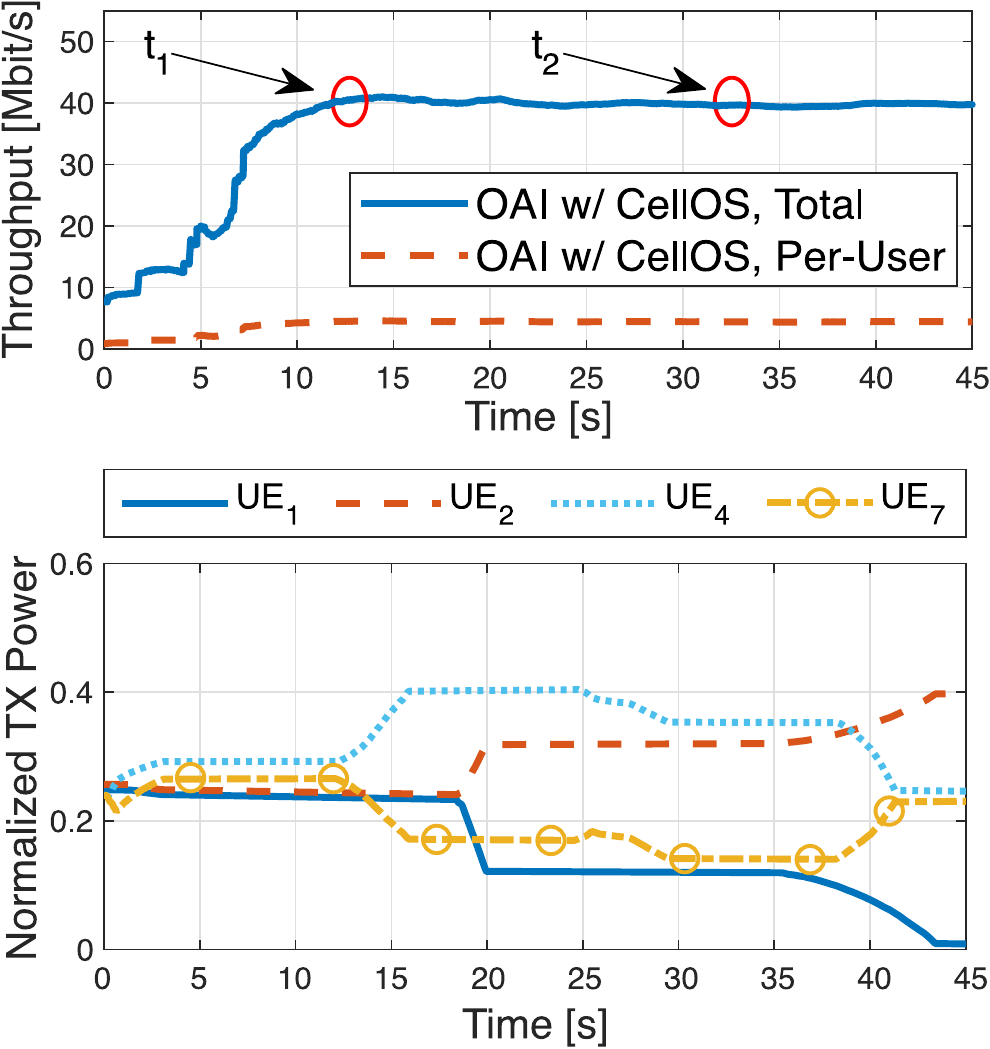}}\hspace{5pt}%
    \end{center}
    \caption{Throughput maximization in the high interference scenario on the OAI-based prototype.}
    \label{fig:maxRate}
\end{minipage}
\begin{minipage}{0.25\textwidth}
    \begin{center}
        \includegraphics[height=4.3cm]{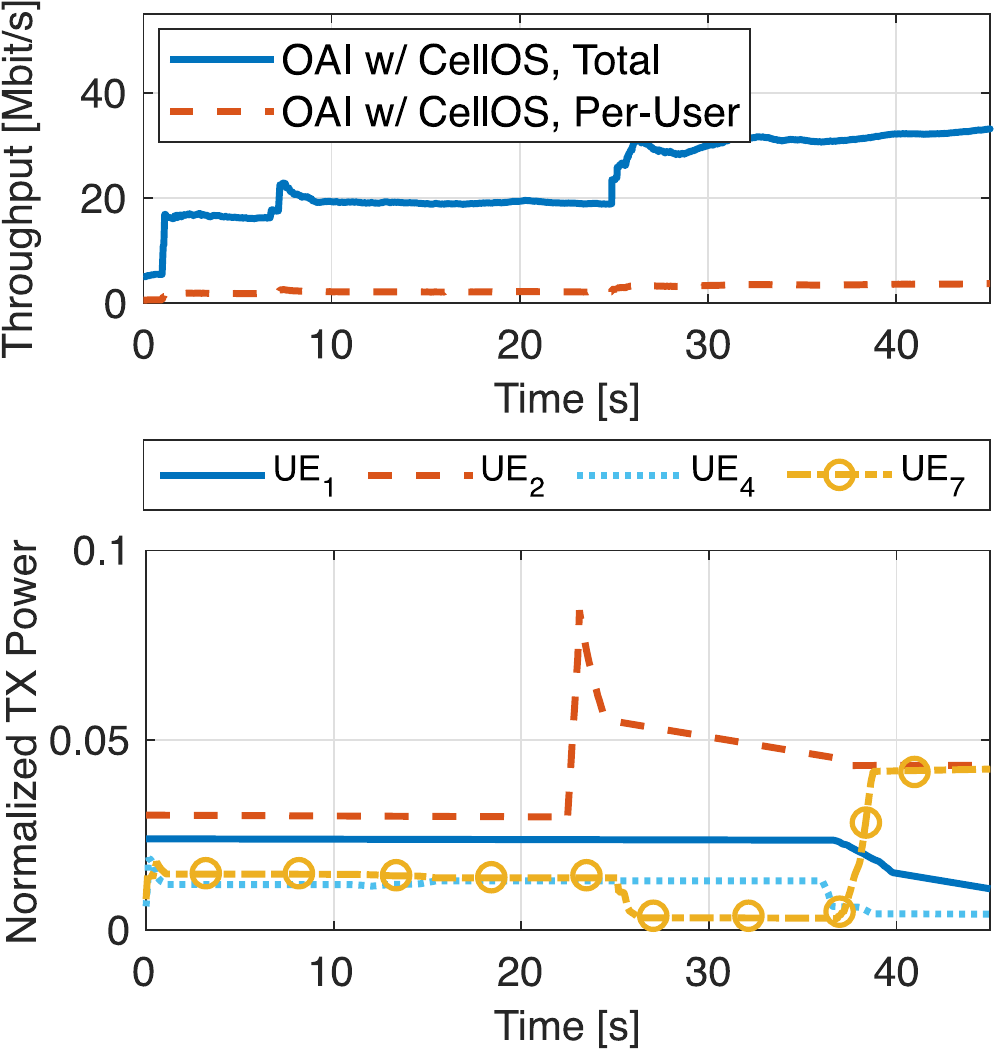}
        \caption{Power minimization in the high interference scenario on the OAI-based prototype.}
        \label{fig:high_interference_min_power_throughput}
    \end{center}
\end{minipage}
\end{minipage}
\end{center}
\end{figure*}

\cellos has been evaluated against the metrics of \sec{metrics} in a variety of network configurations (i.e., high and low interference, with and without network slicing), and on different testbeds, including a lab bench setup, the Arena testbed~\cite{bertizzolo2019arena}, and the \powderrenew \acs{pawr} 5G platform~\cite{pawr, powder}.

To fully appreciate the effects of the automatic optimization procedures introduced by \cellos, we consider a cellular network implemented through OAI and srsLTE and we compare the achieved network performance with and without \cellos.
Moreover, we also compare the performance achieved by state-of-the-art scheduling algorithm commonly used in commercial cellular networks, \textit{i.e., proportional fairness, greedy, and round-robin,} to that achieved by \cellos-managed networks.
A summary of our experimental setup is shown in Table~\ref{tab:experiment_setup}.

\subsubsection{High Interference Scenario}
\label{sec:high_interference}

\fig{maxRate} presents results obtained when optimizing throughput (network control objective of \textit{max(rate)}) in the high interference scenario in \fig{high_interference_scenario}.
We start by evaluating the throughput gains brought to \ac{oai} by \cellos zero-touch approach.
Average total and per-user throughput are shown in \fig{high_interference_max_rate_9ues}.
We observe that \cellos brings significant benefits to the network performance, with improvements as high as $75\%$ ($63\%$ on average).
This is because of the interplay between the optimized per-user power control and scheduling determined by \cellos and executed locally by the Softwarized RAN.
Indeed, \cellos automatic optimization procedures allow the \enbs to serve \ues with an optimized resource allocation and power-controlled signals, which significantly reduces the inter-cell interference while guaranteeing a minimum rate to \ues.
To provide further insights on the resource allocation procedures automatically executed by each \enb, we investigated the network throughput, and power and \acp{prb} allocated to the users during an experiment run of the \textit{max(rate)} solution program (Figures~\ref{fig:high_interference_max_rate_prb} and~\ref{fig:high_interference_max_rate_throughput}, respectively).
For clarity, only the power for four users is shown.
%
%
As time progresses, the throughput (both total and per-user) plateaus out to a stable value, which is a consequence of local optimality of the solution program that successfully limits interference.
Power is changed for the individual user in time, also responding to optimization requirements and reflecting current network conditions.
\fig{high_interference_max_rate_prb} depicts the \acp{prb} allocated to \ues
at time instants $t_1$ and $t_2$ of \fig{high_interference_max_rate_throughput}.
We observe that the \enbs{} adapt the \ac{prb} allocation in \textit{real-time} to satisfy user requests while achieving the set network objective.
In fact, time slots with unassigned \acp{prb} may even occur, without compromising the \enb ability of satisfying its subscribers requirements.

To show that different network control objectives produce different results, we investigate throughput and power determined by \cellos for power minimization (control objective of \textit{min(power)}), while guaranteeing a minimum per-user data rate of $1\:\mathrm{Mbit/s}$ (\fig{high_interference_min_power_throughput}).
As expected, the achieved throughput is lower than that of the \textit{max(rate)} control program (\fig{high_interference_max_rate_throughput}).
This is due to the normalized transmission power of the \enbs being remarkably lower than that in \fig{high_interference_max_rate_throughput} (up to one order of magnitude).
We notice, though that \ues achieve an average throughput of $2.63\:\mathrm{Mbit/s}$, which satisfies the constraint on their minimum rate.
%
%
%

\begin{figure}[ht]
    \begin{center}  	
        \subcaptionbox{\label{fig:thr_varying_ues}Throughput.}{\includegraphics[width=0.3\columnwidth]{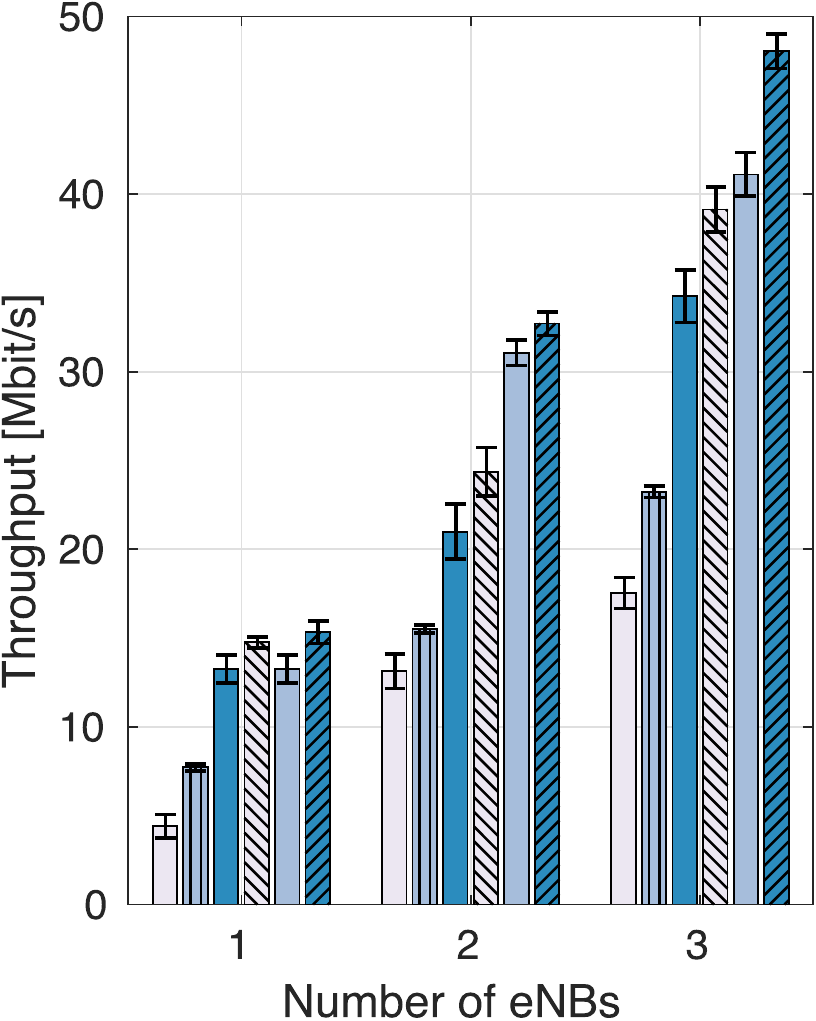}}\hspace{0.019\columnwidth}%
        \subcaptionbox{\label{fig:ee_varying_ues}Energy efficiency.}{\includegraphics[width=0.3\columnwidth]{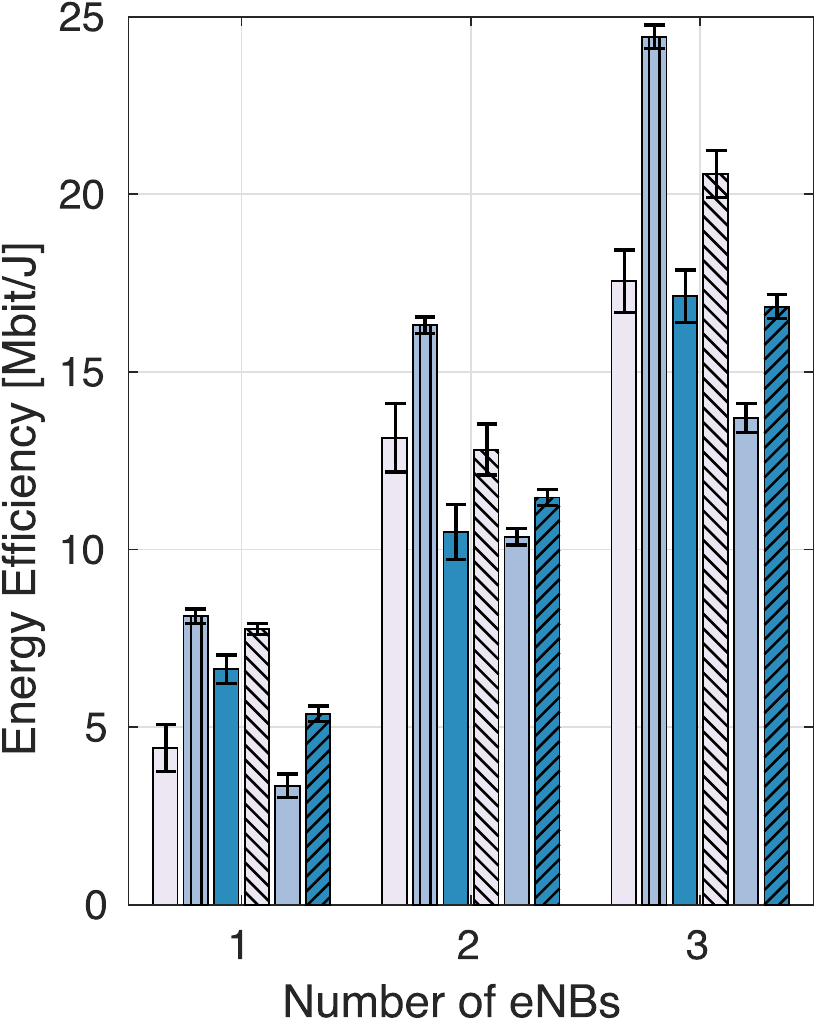}}\hspace{0.019\columnwidth}%
        \subcaptionbox{\label{fig:fair_varying_ues}Fairness.}{\includegraphics[width=0.3\columnwidth]{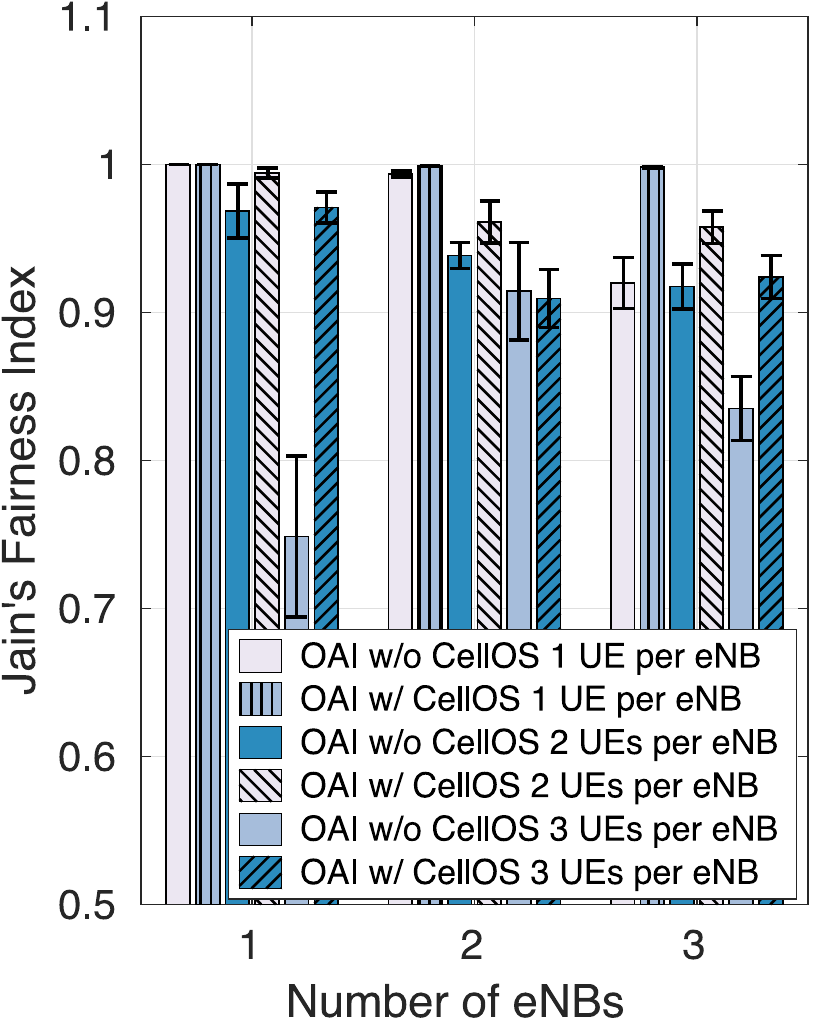}}
    \end{center}
    \caption{Sum-log-rate maximization in the high interference scenario on the OAI-based prototype w/ and w/o \cellos.}
    \label{fig:bar_plots}
\end{figure}

The next set of experiments concerns the performance of OAI with and without \cellos in scenarios with varying number of \enbs{} and UEs.
The network control objective requires to maximize throughput while explicitly accounting for fairness, namely, is set to \textit{max(sum(log(rate)))}.
Scenarios with one base station consider only~\mathenb{3}, while Scenarios with two base stations concern~\mathenb{2} and~\mathenb{3}, i.e., the base stations with overlapping cells (see \fig{high_interference_scenario}).
Results concerning sum throughput, energy efficiency and fairness are shown in \fig{bar_plots}.

The throughput comparison is shown in \fig{thr_varying_ues}, where we can see that \ac{oai} with \cellos always outperforms \ac{oai} without \cellos.
%
%
%
In \fig{ee_varying_ues}, we evaluate energy efficiency, pivotal in large-scale networks~\cite{3gpp.32.972}.
As expected, since our framework achieves a higher throughput with a lower power expenditure, the network is more energy efficient when managed by \cellos.
System fairness is shown in \fig{fair_varying_ues}.
We notice that, in general, \cellos improves user fairness, 
with increases
up to~$29\%$.
Improvements are more evident in scenarios with higher number of \enbs and \ues, as optimization techniques are more effective in those more dense scenarios with higher interference.
Specifically, since in these scenarios suboptimal algorithm solutions generate inefficient resource allocation policies, optimal ones are required the most.
Indeed, \cellos optimized resource allocation, and its ability to fine-tune the power directed to the served \ues allows the base stations to contain the interference directed to other \enbs, thus increasing the network performance.
%

\subsubsection{Low Interference Scenario}

\begin{figure}[ht]
    \centering
    \includegraphics[width=0.65\columnwidth]{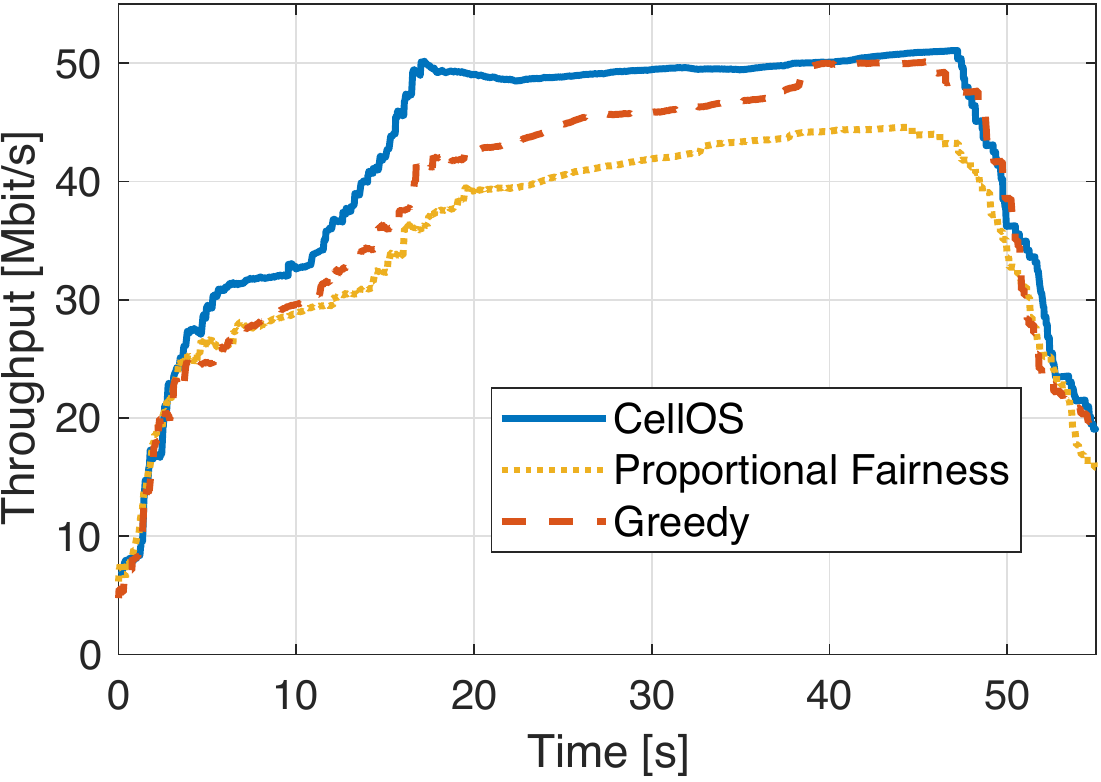}
    \caption{Rate maximization in the low interference scenario: OAI w/ \cellos vs.\ OAI w/ \textit{proportional fairness} \cite{bu2006generalized} and OAI w/ \textit{greedy} \cite{wu2007scheduling} scheduling policies.}
    \label{fig:low_interference_scheduling}
\end{figure}

These experiments concern~3~\enbs and 9~\ues in low interference conditions (\fig{low_interference_scenario}).
%
%
\begin{figure}[ht]
    \centering
    \includegraphics[width=0.5\columnwidth]{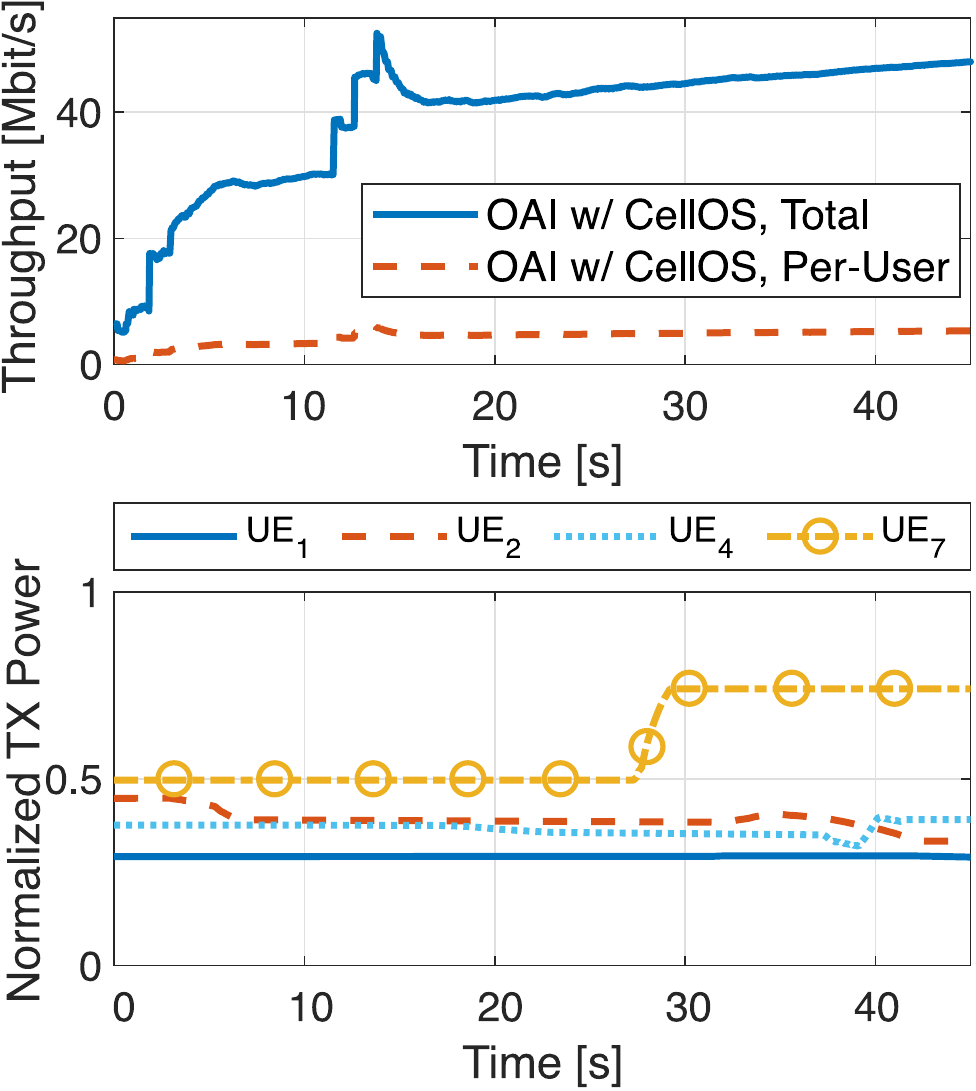}
    \caption{Sum-log-rate maximization in the low interference scenario on the OAI-based prototype w/ \cellos.}
    \label{fig:low_interference_sumlog_rate_throughput}
\end{figure}
Results on throughput and on the allocated normalized power are shown in \fig{low_interference_sumlog_rate_throughput}.
In this scenario \cellos is required to optimize the network control objective \textit{max(sum(log(rate)))}.
As expected, performance is better than in the high interference scenario because of the lower interference level,
that allows the \enbs{} to use higher power without disrupting each other transmissions.
In \fig{low_interference_scheduling}, we compare \cellos rate maximization with two well-known state-of-the-art scheduling algorithms: The \textit{proportional fairness} algorithm, that is the \textit{de facto standard} in cellular networks~\cite{bu2006generalized, margolies2016exploiting}, and the \textit{greedy} algorithm~\cite{wu2007scheduling}.
We notice that \cellos outperforms the proportional fairness algorithm because of this overarching optimization approach to network management.
The greedy approach, instead, obtains throughput levels similar to those of \cellos, albeit with a significant delay.
Indeed, because of its optimized \ac{mac}-layer procedures, which allow the network base stations to mindfully allocate resources to the served \ues, \cellos achieves said throughput level after only few seconds from the system start and maintains it until the \ues finish downloading data.

\subsubsection{Network Slicing}
\label{sec:slicing}

This set of experiments concerns 3~\enbs instantiated on the \acp{usrp} X310 of the Arena testbed~\cite{bertizzolo2019arena} through srsLTE.
The \enbs serve 9~\ac{cots} \ues.
The antennas of the \acp{usrp} are hung off the ceiling of a $208.1\:\mathrm{m^2}$ office space.

We target a scenario in which multiple \acp{mvno} lease infrastructure resources from an \ac{ip}.
The \ac{ip}, which owns the physical equipment (e.g., the base stations), allocates slices of the network to \acp{mvno} following, for instance, the approach described in~\cite{doro2020sledge}.
Since \acp{mvno} act independently from one another, with different subscribers and requirements (e.g., quality of service), they may need to optimize different control programs on their slice of the network.
Considering this, and cognizant of current 5G cellular networks trends, we designed \cellos to handle different network slicing configurations.

\begin{figure*}[ht]
    \begin{center}
    \centerline{
        \subcaptionbox{\label{fig:slice_max_rate}}{\includegraphics[height=5cm]{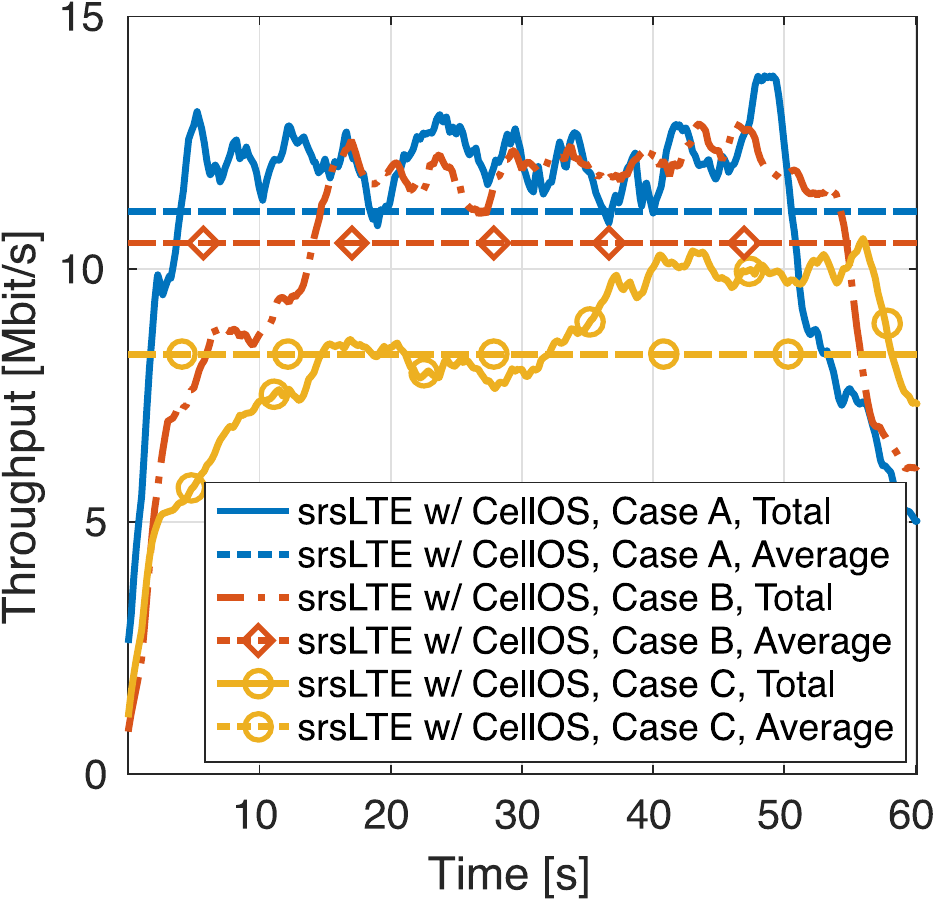}}\hspace{15pt}%
        \subcaptionbox{\label{fig:slice_min_power}}{\includegraphics[height=5cm]{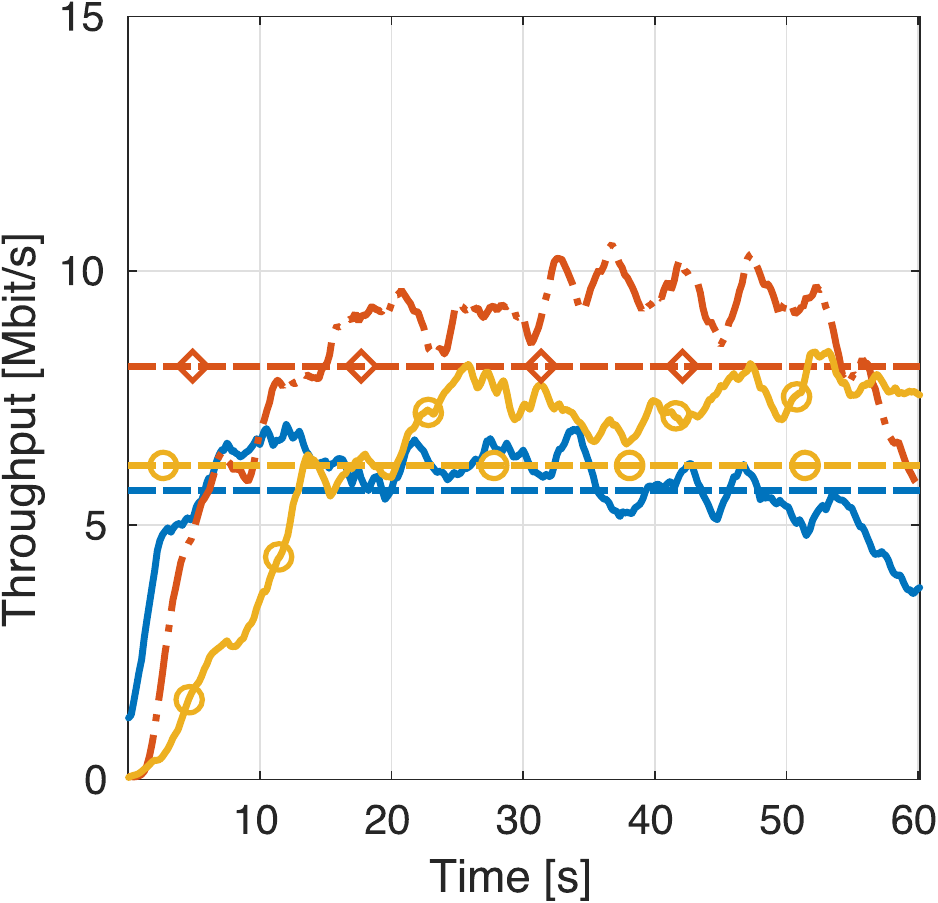}}\hspace{20pt}%
        \subcaptionbox{\label{fig:slice_allocation}}{\includegraphics[trim=0 22 0 0, clip, height=5cm]{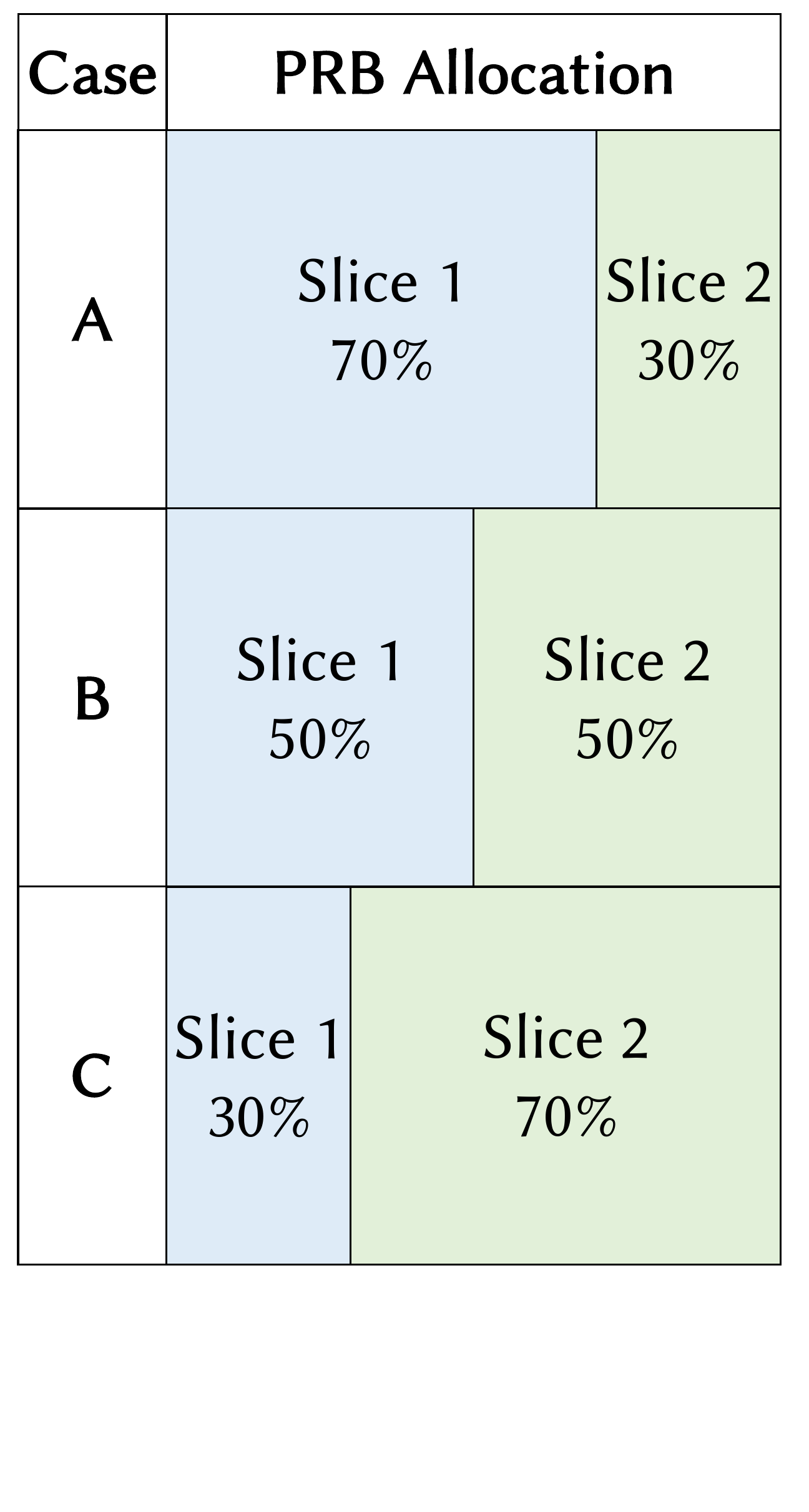}}
        }
    \end{center}
    \caption{Optimization of different control programs on different slices on the srsLTE-based prototype instantiated on the Arena testbed~\cite{bertizzolo2019arena}: (a) Throughput of \slice{1} (\textit{max(rate)}); (b) throughput of \slice{2} (\textit{min(power)}); (c) \acs{prb} allocation.}
    \label{fig:slicing}
\end{figure*}

\fig{slicing} showcases the unique ability of \cellos in implementing different control strategies for different network slices, \textit{simultaneously optimizing different control programs on different network slices,} namely, \slice{1} and \slice{2}, on each \enb.
Specifically, \slice{1}, which is allocated to \ac{mvno}~1, aims at maximizing the network throughput, while \slice{2}, allocated to \ac{mvno}~2, minimizes the power consumption.
%
%
The network sum and average throughput achieved by this per-slice behavior are shown in \fig{slicing}.
In our experiments, the two slices were allocated different percentages of the available \acp{prb} (see \fig{slice_allocation}): First 70\% to \slice{1} and 30\% to \slice{2} (Case A of \fig{slicing}), then 50\% to each slice (Case B), and finally a 30\%--70\% allocation was used (Case C).
%
\fig{slice_max_rate} shows the throughput of \slice{1} in the three cases. \fig{slice_min_power} presents that of \slice{2}.
As expected, the throughput of the \textit{max(rate)} control program instantiated by \ac{mvno}~1 on \slice{1} increases with the resources allocated to the slice.
On the contrary, the throughput performance of the \textit{min(power)} control program instantiated by \ac{mvno}~2 on \slice{2} does not increase with the resources allocated to the slice.
All three configurations of \fig{slice_min_power} converge toward $7\:\mathrm{Mbit/s}$.
This is due to the fact that this control problem aims at reaching the minimum per-user rate constraint set by the \telco without consuming all available network resources.
By looking at \fig{slicing}, we notice that \cellos managed to independently optimize different control problems on different slices of the network (\textit{max(rate} on \slice{1}, and \textit{min(power)} on \slice{2}).
This demonstrates that \cellos provides softwarized \acp{mvno} with independent control of all resources in their leased network slice while sharing the same physical network infrastructure.

\subsubsection{CellOS Scalability}
\label{sec:scalability}

In this section, we evaluate the scalability of \cellos in terms of time and operations required by the controller to generate distributed solution programs, and by the \acp{ree} to solve them. Finally, we compare the overhead generated by \cellos \acp{ree} to that of state-of-the-art solutions, such as FlexRAN~\cite{foukas2016flexran} and Orion~\cite{foukas2017orion}.
\begin{figure}[ht]
    \centering
    \includegraphics[width=0.75\columnwidth]{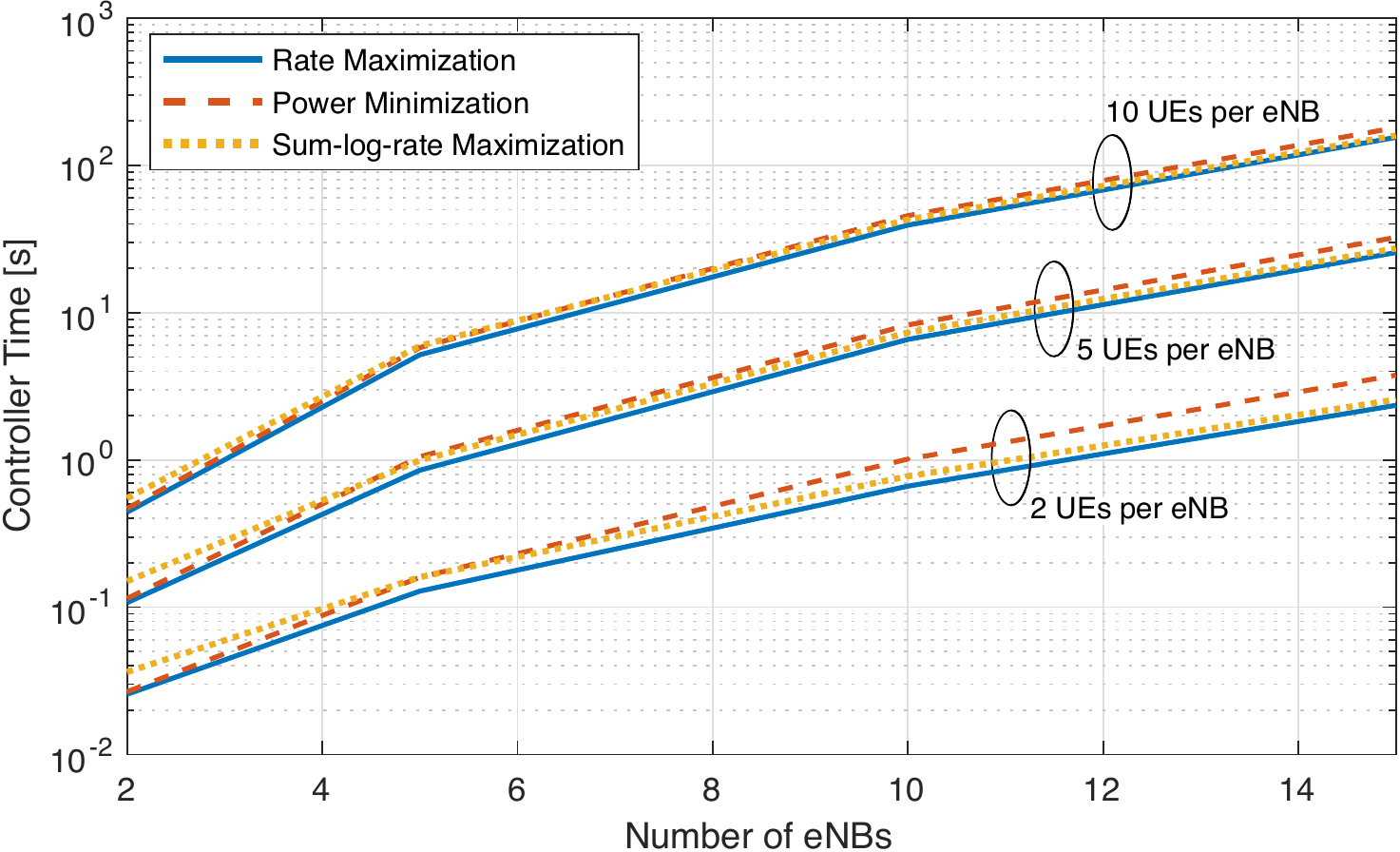}
    \caption{Scalability of \cellos controller operations as a function of the number of \enbs, \ues and for different network control problems.}
    \label{fig:scalability}
\end{figure}
The results presented in this section have been obtained by executing \cellos on a single CPU of a Dell EMC PowerEdge R340 server of the Arena testbed~\cite{bertizzolo2019arena}. The server is equipped with an Intel Xeon E-2146G processor with $3.5\:\mathrm{GHz}$ base frequency
and $32\:\mathrm{GB}$ DDR4-2666 RAM.

\fig{scalability} shows the time needed by \cellos controller to generate the distributed solution programs starting from the \telco directives as a function of the number of network \enbs, \ues, and for different network control problems.
This
includes the time to perform: (i) The problem definition procedures, which interpret the \telco high level directives; (ii) the generation of the centralized version of the problem based on an abstraction of the network, and (iii) the problem decomposition operations, which divide the centralized problem into sub-problems to be distributively solved by the softwarized \ac{ran}.
We notice that, even though the computation time
increases with the number of users and base stations, these operations are executed once per control problem. Also, recall that the generated problems utilize symbolic placeholders and do not require knowledge of real-time parameters. For this reason, all operations can be performed offline, and computation times are thus negligible if compared to the typical service times of cellular networks. 

\begin{figure}[ht]
    \centering
    \includegraphics[width=0.75\columnwidth]{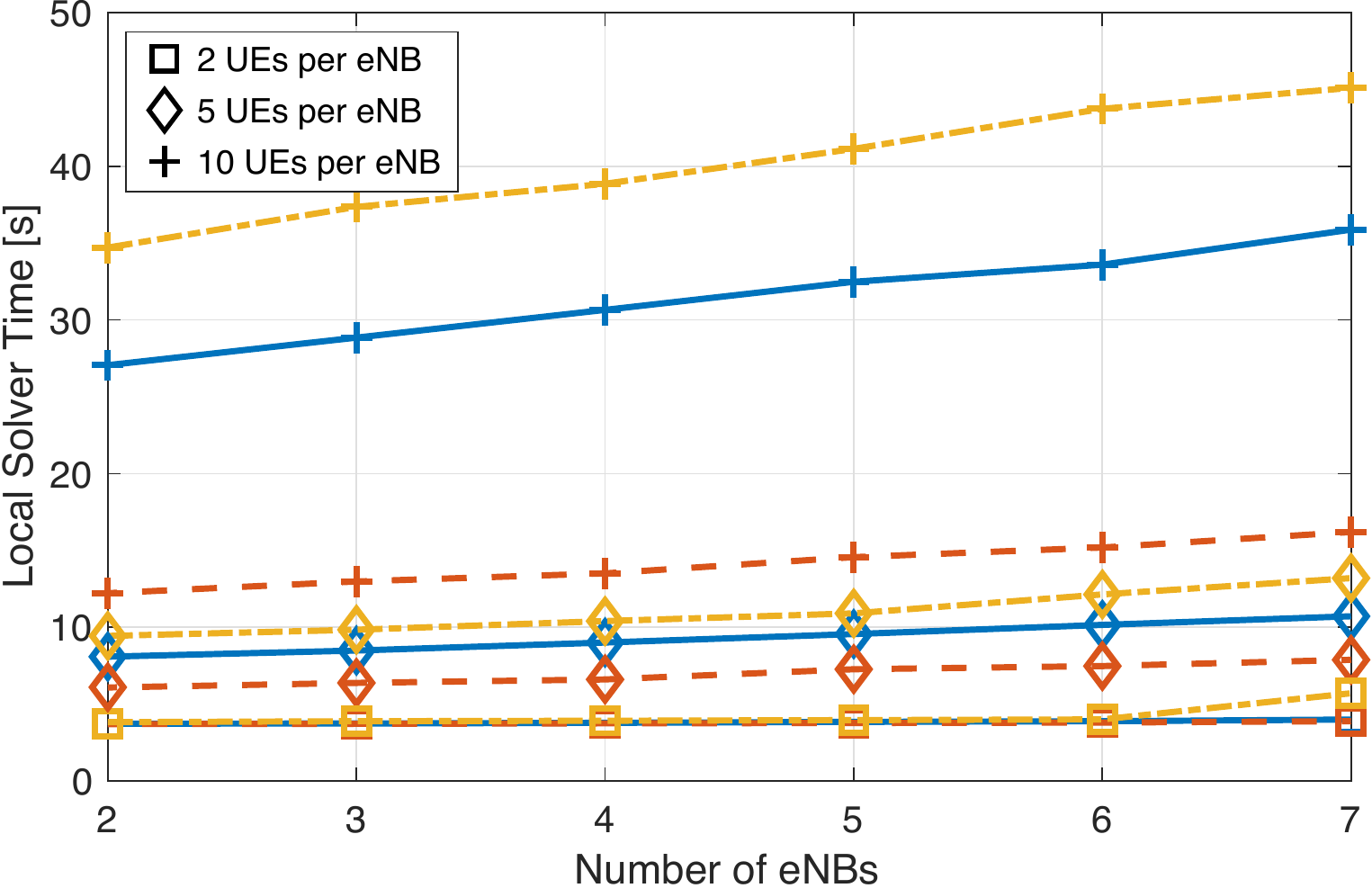}
    \caption{Scalability of \cellos local solver operations as a function of the number of \enbs, \ues and for different network control problems: (i) Rate maximization (solid lines); (ii) sum-log-rate maximization (dot-dashed lines), and (iii) power minimization (dashed lines).}
    \label{fig:local_solver}
\end{figure}

\fig{local_solver} shows the time needed by \cellos \acp{ree} to solve the distributed problems automatically generated by the controller (\sec{framework}) for different numbers of base stations and \ues in the network.
Different control problems require different solution times.
For instance, the power minimization problem, whose objective function is a linear function in the transmission power variables, is solved more rapidly than the rate and sum-log-rate maximization problems, whose utility functions are non-linear because of logarithmic and fractional terms, which increase the problem complexity. As a consequence, the execution time of each problem strongly depends on the complexity of the underlying objective function to be optimized.
It is worth noticing that the times of both Figures~\ref{fig:scalability} and~\ref{fig:local_solver} can be considerably reduced if executed on high-performance equipment, as the one typically used in commercial cellular network deployments.

\begin{figure}[ht]
    \centering
    \includegraphics[width=0.75\columnwidth]{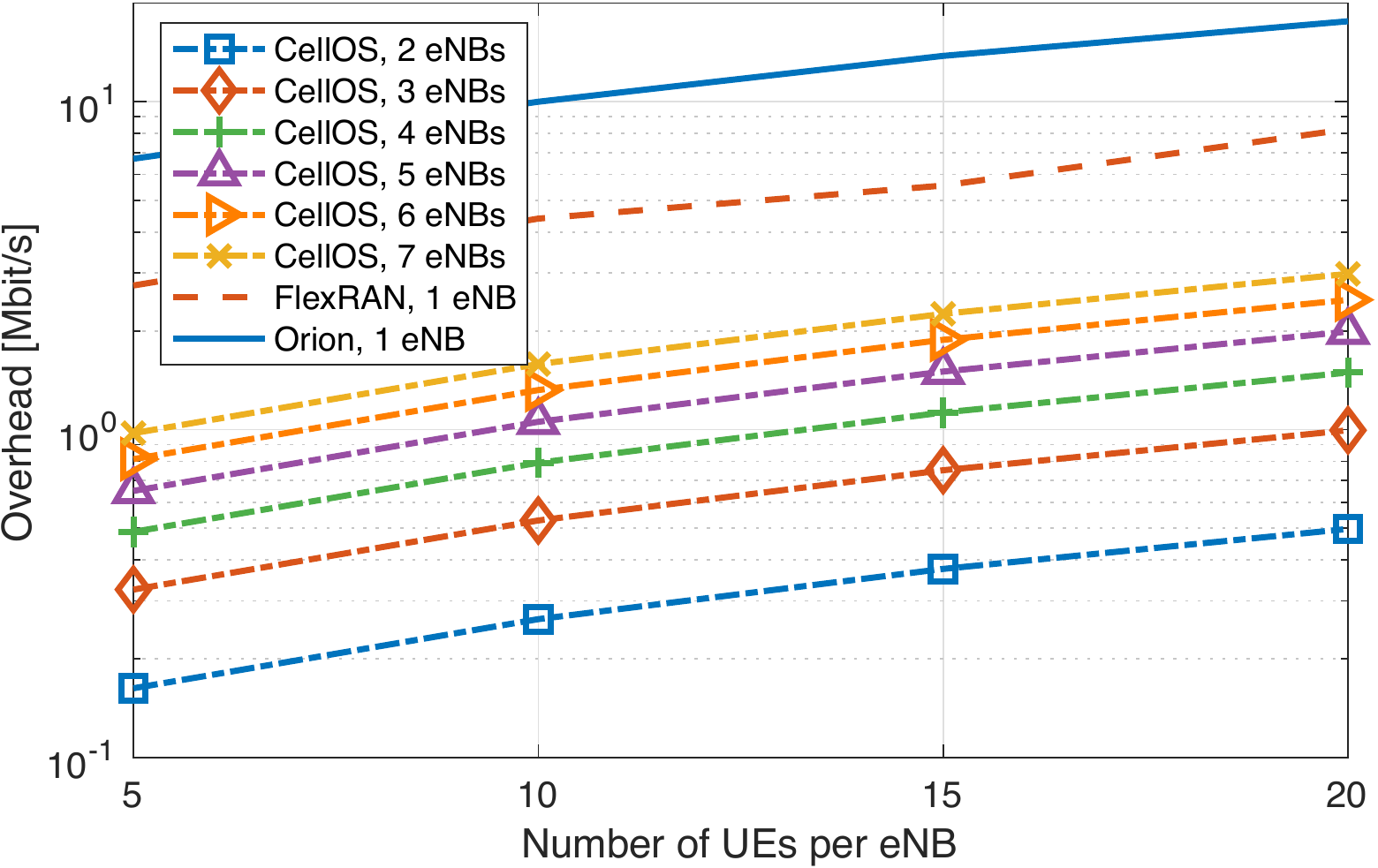}
    \caption{Signaling overhead: \cellos vs.\ FlexRAN~\cite[Figure~7]{foukas2016flexran} and Orion~\cite[Figure~13a]{foukas2017orion}.}
    \label{fig:overhead}
\end{figure}

The signaling overhead generated by each \cellos \ac{ree} is evaluated in \fig{overhead} against that generated by other well-established software-defined cellular control frameworks such as Flex\-RAN~\cite{foukas2016flexran} and Orion~\cite{foukas2017orion}.
Since \cellos executes the optimization problems locally at each \ac{ree}, its overhead stems from the \acp{ree} exchanging $|\mathcal{U}| \, (|\mathcal{N}| \!+\! 1)$ optimization variables and Lagrangian multipliers.
These are the only information required to converge to a distributed problem solution (\sec{infrastructure}).
These variables are represented by real numbers encoded as 32-bit floating point numbers.
%
\fig{overhead} shows that the signaling overhead generated by \cellos \acp{ree} is significantly lower than that of prevailing state-of-the-art centralized approaches. 
Even when managing a single network base station, as it is the case of \fig{overhead}, previous approaches must exchange a massive amount of local information with the central controller, thus generating large signaling and latency.

\subsubsection{Experiment of \powderrenew \acs{pawr} Platform}
\label{sec:powder}

\begin{figure}[htb]
    \begin{center}
    \subcaptionbox{\label{fig:powder_thr}}{\includegraphics[height=3.85cm]{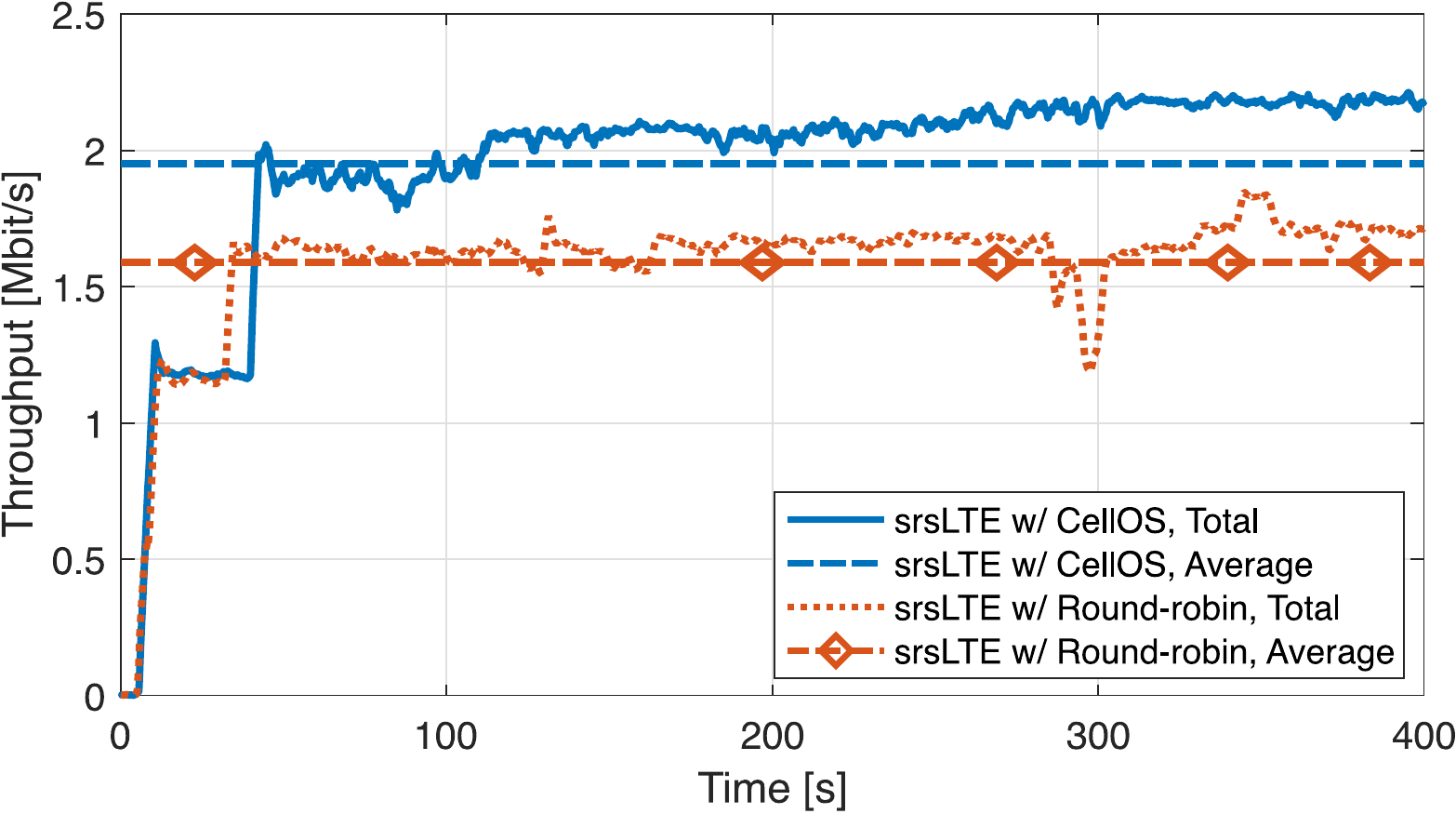}}\\\vspace{10pt}%
    \subcaptionbox{\label{fig:powder_map}}{\includegraphics[height=3.85cm]{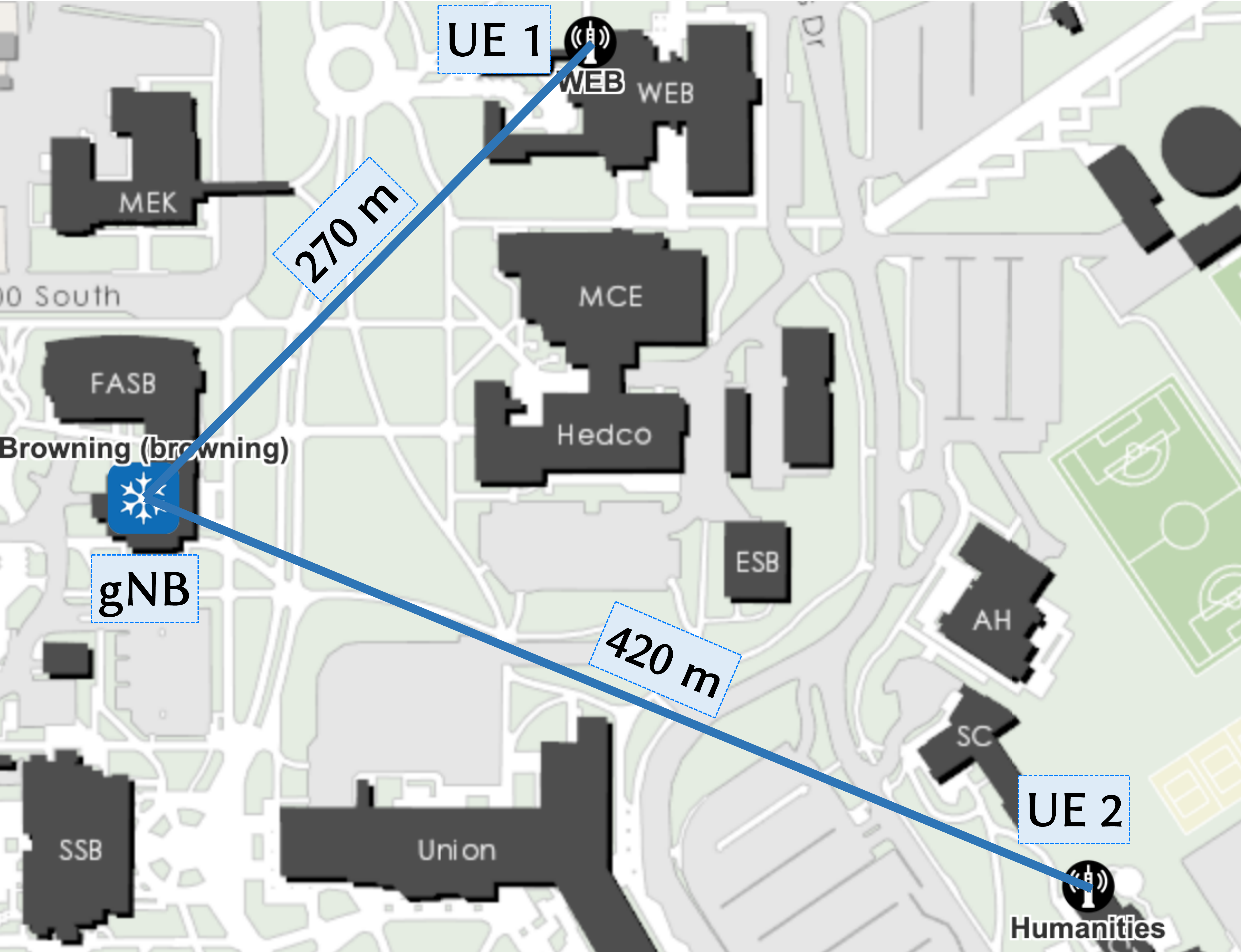}}
    \end{center}
    \caption{Long-range experiments on the \powderrenew \acs{pawr} platform~\cite{pawr, powder}: (a)~srsLTE w/ \cellos rate maximization vs.\ srsLTE w/ \textit{round-robin}; (b)~long-range experiment area.}
    \label{fig:powder}
\end{figure}

We demonstrate the platform- and \acs{ran}-independence of \cellos by running  \textit{long-range} experiments on one of the \ac{pawr} wireless platforms~\cite{pawr}.
Specifically, we leverage \powderrenew~\cite{powder} and the 5G implementation of srsLTE to deploy a NR~\gnb and 2~\ues in an authentic outdoor wireless environment.
The \gnb employs a \ac{usrp} X310 located on the rooftop of a $28.75\:\mathrm{m}$-tall building, while we use ground-level \acp{usrp} B210 as \ues.
The \gnb utilizes a reduced channel bandwidth of 15~\acp{prb} (corresponding to $3\:\mathrm{MHz}$) to reach the two \ues distant $270\:\mathrm{m}$ and $420\:\mathrm{m}$, respectively (see \fig{powder_map}).
In this case, the \ues download a file from a local server for $400\:\mathrm{s}$.

\fig{powder_thr} shows the throughput gains achievable by running \cellos rate maximization on top of srsLTE, which uses a \textit{round-robin} scheduler when instantiated without \cellos.
Albeit the reduced bandwidth and increased \gnb-\ues distance result in a lower total throughput than that of the previous experiments, we notice that \cellos significantly improves the network performance because of its zero-touch approach to optimization, which allows to optimize the resources allocated to the \ues, and bring gains as high as $86\%$ ($23\%$ on average).
To the best of our knowledge, this is the first demonstration of \textit{zero-touch optimization} on a long-range open-source 5G testbed.
Such instantiation gives evidence of the potential of the \textit{softwarized Open RAN} approach cellular networks are moving toward.

\section{Related Work}
\label{sec:related}

Recent years have heralded \ac{sdn} as the technology that would inherently endow the \textit{monolithic} Internet architecture with much needed \textit{flexibility}.
The largest part of \ac{sdn} work focuses on the \textit{programmability} of wired networks, with few works exploring
scenarios comprising wireless devices~\cite{bansal2012openradio, gudipati2013softran, li2012cellsdn, bradai2015cellular, thembelihle2017softwarization, foukas2016flexran, guan2018wnos}.
To the best of our knowledge, there is no solution aimed at integrating a zero-touch, flexible, and dynamic optimization framework to the fabric of cellular networks.
%
Therefore,
this section reviews
SDN-based solutions for wireless networking.

Guan et al.\ proposed WNOS, a wireless network operating system featuring network virtualization and distributed solution of optimization problems~\cite{guan2018wnos}.
Although this work is the most similar to ours, it only focuses on infrastructure-less ad hoc networks with static nodes.
For this reason, it is not suitable to handle mobile and dynamic cellular scenarios.
An effort to explicitly take mobility into account is made by Bertizzolo et al.\ with SwarmControl, a distributed control framework for the self-optimization of drone networks~\cite{bertizzolo2020swarmcontrol}.

\ac{onap} and O-RAN are two infrastructure-oriented automation platforms with the ambition of ``orchestrating'' many network functions~\cite{linux2018onap, oran2018oran}.
%
They offer \acp{to} network abstractions to specify system details and traffic policies.
However, optimization policies and algorithms must be explicitly programmed.  

Adaptations of the SDN paradigm to cellular networks have been proposed by Li et al.~(CellSDN~\cite{li2012cellsdn}), Bernardos et al. (SDWN~\cite{bernardos2014architecture}), and by Bradai et al.~(CSDN~\cite{bradai2015cellular}).
CellSDN proposes a control-oriented operating system
%
focused on cellular network management and subscriber policies rather than on performance optimization.
%
%
Works like SDWN and CSDN, instead, describe general frameworks to optimize network utilization and performance leveraging edge network information.

Few works have addressed the interplay between the SDN architecture and that of networks including \ac{lte} explicitly.
Gudipati et al.\ envision SoftRAN as an abstraction of all \enbs{} in a geographical area as a single virtual base station to perform operations including 
metrics optimization~\cite{gudipati2013softran}.
%
%
This centralized approach,
however,
can hardly address heterogeneous optimization problems in the dense, flexible and rapidly growing architecture of 5G cellular networks.
%
Foukas et al.\ propose FlexRAN~\cite{foukas2016flexran} and Orion~\cite{foukas2017orion} as centralized controllers coordinating various \ac{lte} agents, and supporting network slicing, respectively.
These systems, though, neglect optimization, and their centralized nature may result in limited scalability and reduce the performance in dense scenarios.
Finally, OpenRadio, by Bansal et al., develops a programmable wireless data plane 
providing
programming interfaces on \ac{phy}
and \ac{mac} layers~\cite{bansal2012openradio}.
Optimization, however, is left to the wits of the TO.

Finally, we notice that all the mentioned solutions for cellular networks propose programmable protocol stack implementations where the optimization procedures need to be \textit{manually designed} and there is no way to perform them \textit{dynamically} or  \textit{automatically}.


\section{Conclusions}
\label{sec:conclusions}

We presented \cellos, the first zero-touch optimization and management framework
for next-generation cellular open \acp{ran}.
%
\cellos enables \acp{to} to automatically optimize the \textit{network behavior} through high-level directives without requiring knowledge of optimization theory or of network specifics.
\cellos \textit{automatically} generates distributed solution programs to be run at the base stations to simultaneously optimize heterogeneous objectives on different network slices.
We prototyped \cellos by using the \ac{lte}-compliant \acl{oai} and srsLTE software, and demonstrated its capabilities through a experimental campaign under varying indoor settings, characterized by different interference conditions and heterogeneous devices.
Results indicate that \cellos remarkably improves key performance metrics when compared with existing solutions, including throughput (up to $86\%$), energy efficiency (up to $84\%$), and user fairness (up to $29\%$).
Finally, we evaluated \cellos in the outdoor environment of the \powderrenew \ac{pawr} 5G platform, providing long-range links.
Results from those experiments confirm the effectiveness of \cellos in obtaining superior performance and indicate a new way of managing and optimizing softwarized cellular networks.

\balance

\footnotesize
\bibliographystyle{IEEEtran}
\bibliography{biblio}

\begin{IEEEbiography}
[{\includegraphics[width=1in,height=1.25in,keepaspectratio]{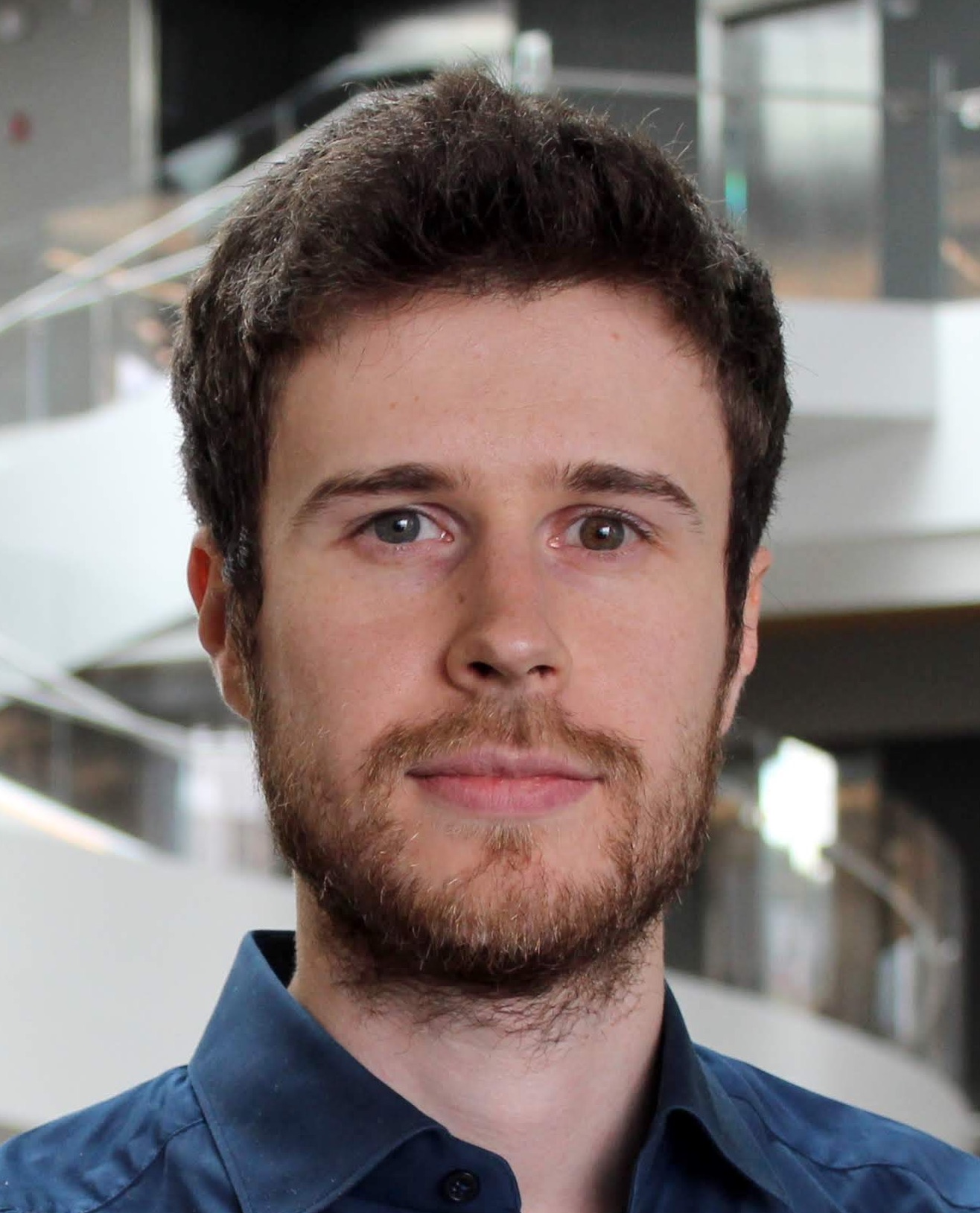}}]{Leonardo Bonati}
received his B.S. in Information Engineering and his M.S. in Telecommunication Engineering from University of Padova, Italy in 2014 and 2016, respectively. He is currently pursuing a Ph.D.\ degree in Computer Engineering at Northeastern University, MA, USA. His research interests focus on 5G cellular networks, network slicing, software-defined networking for wireless networks, and unmanned aerial vehicles networks.
\end{IEEEbiography}

\vspace{-1cm}

\begin{IEEEbiography}
[{\includegraphics[width=1in,height=1.25in,keepaspectratio]{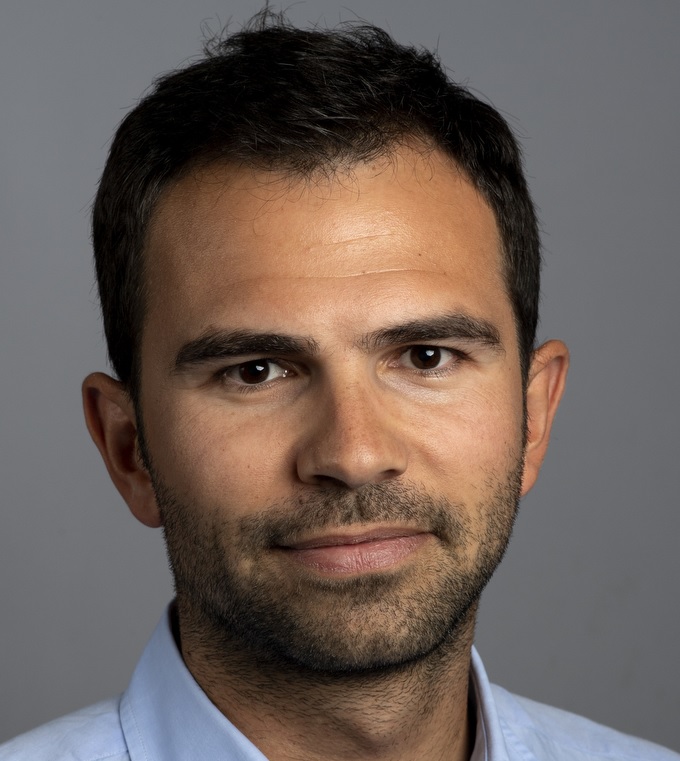}}]{Salvatore D'Oro}
received received his  Ph.D.\ degree from the University of Catania in 2015. He is currently an Associate Research Scientist at Northeastern University. In 2015, 2016 and 2017 he organized the 1st, 2nd and 3rd Workshops on COmpetitive and COoperative Approaches for 5G networks (COCOA). He also served on the Technical Program Committee (TPC) of the IEEE Conference on Standards for Communications and Networking (CSCN'18), Med-Hoc-Net 2018 and the CoCoNet8 workshop at IEEE ICC 2016. He serves on the TPC of Elsevier Computer Communications journal. Dr.\ D'Oro is also a reviewer for major IEEE and ACM journals and conferences. Dr.\ D'Oro's research interests include game-theory, optimization, learning and their applications to telecommunication networks. He is a Member of the IEEE.
\end{IEEEbiography}

\vspace{-1cm}

\begin{IEEEbiography}
[{\includegraphics[width=1in,height=1.25in,keepaspectratio]{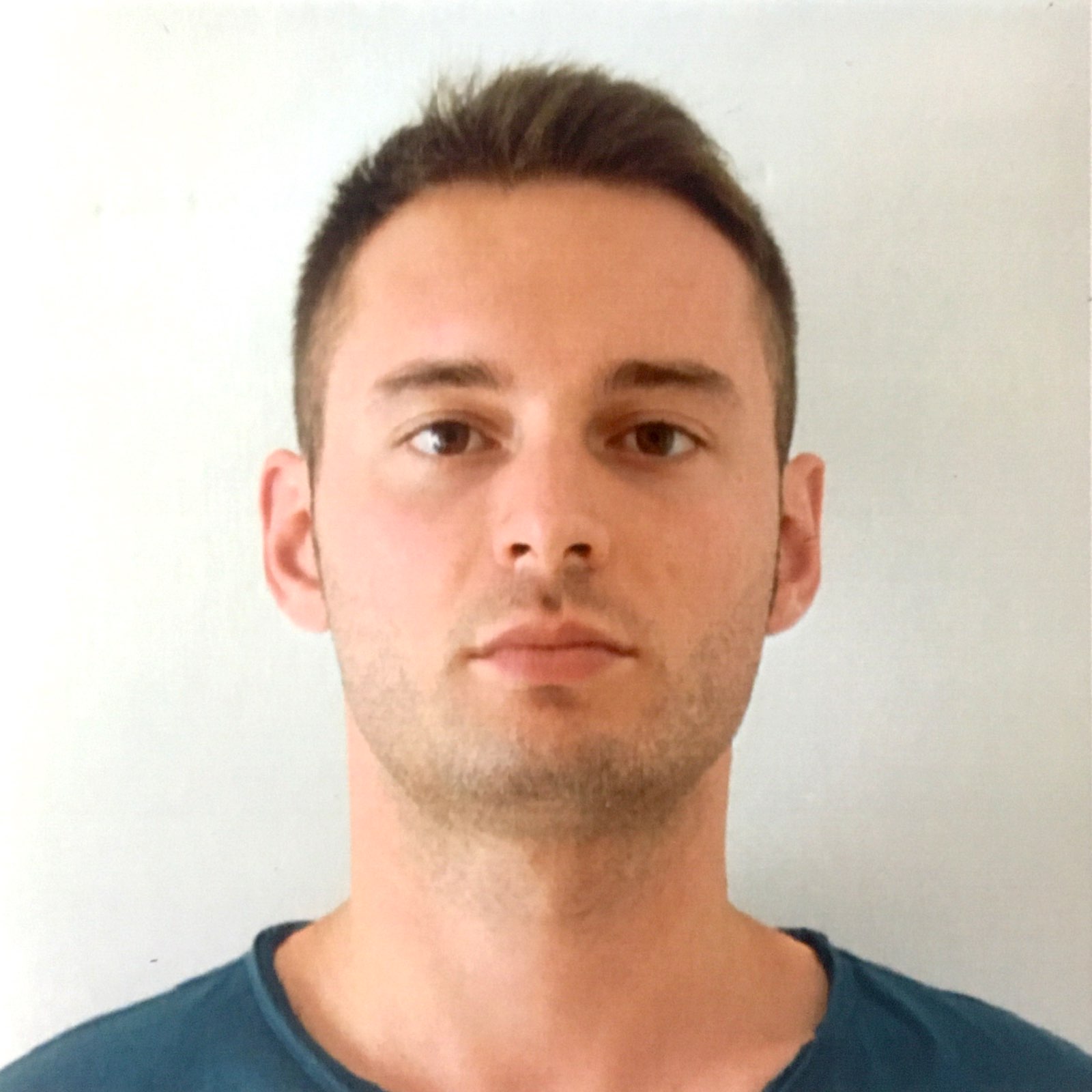}}]{Lorenzo Bertizzolo}
is a candidate for Ph.D.\ in Computer Engineering and research assistant at the Institute for the Wireless Internet of Things at Northeastern University and a collaborator of AT\&T Labs Research, working on the integration of Unmanned Aerial System into the next generations' cellular networks. 
He earned his B.S.\ and his M.S.\ in Computer and Communication Networks Engineering from Politecnico di Torino, Italy in 2014 and 2015, respectively. 
His research focuses on 5G, software-defined networking for wireless networks, distributed optimization, and Unmanned Aerial Networks.
\end{IEEEbiography}

\vspace{-1cm}

\begin{IEEEbiography}
[{\includegraphics[width=1in,height=1.25in,keepaspectratio]{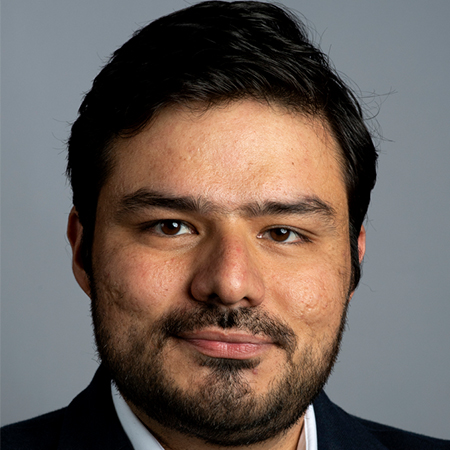}}]{Emrecan Demirors}
is a Research Assistant Professor with the Department of Electrical and Computer Engineering at Northeastern University. He is conducting research at the Wireless Networks and Embedded Systems Laboratory. Previously, he was an Associate Research Scientist with the Department of Electrical and Computer Engineering at Northeastern University, from 2017 to 2019. He received my Ph.D.\ degree in Electrical and Computer Engineering from Northeastern University in 2017, under the supervision of Professor Tommaso Melodia. He had previously received my B.S. and M.S degrees in Electrical and Electronics Engineering from Bilkent University, Ankara, Turkey in 2009 and 2011, respectively, under the supervision of Professor Hayrettin Koymen. From 2010 to 2011, he was a Systems Engineer at Meteksan Defence Industry Inc., Ankara, Turkey.
\end{IEEEbiography}

\vspace{-1cm}

\begin{IEEEbiography}
[{\includegraphics[width=1in,height=1.25in,keepaspectratio]{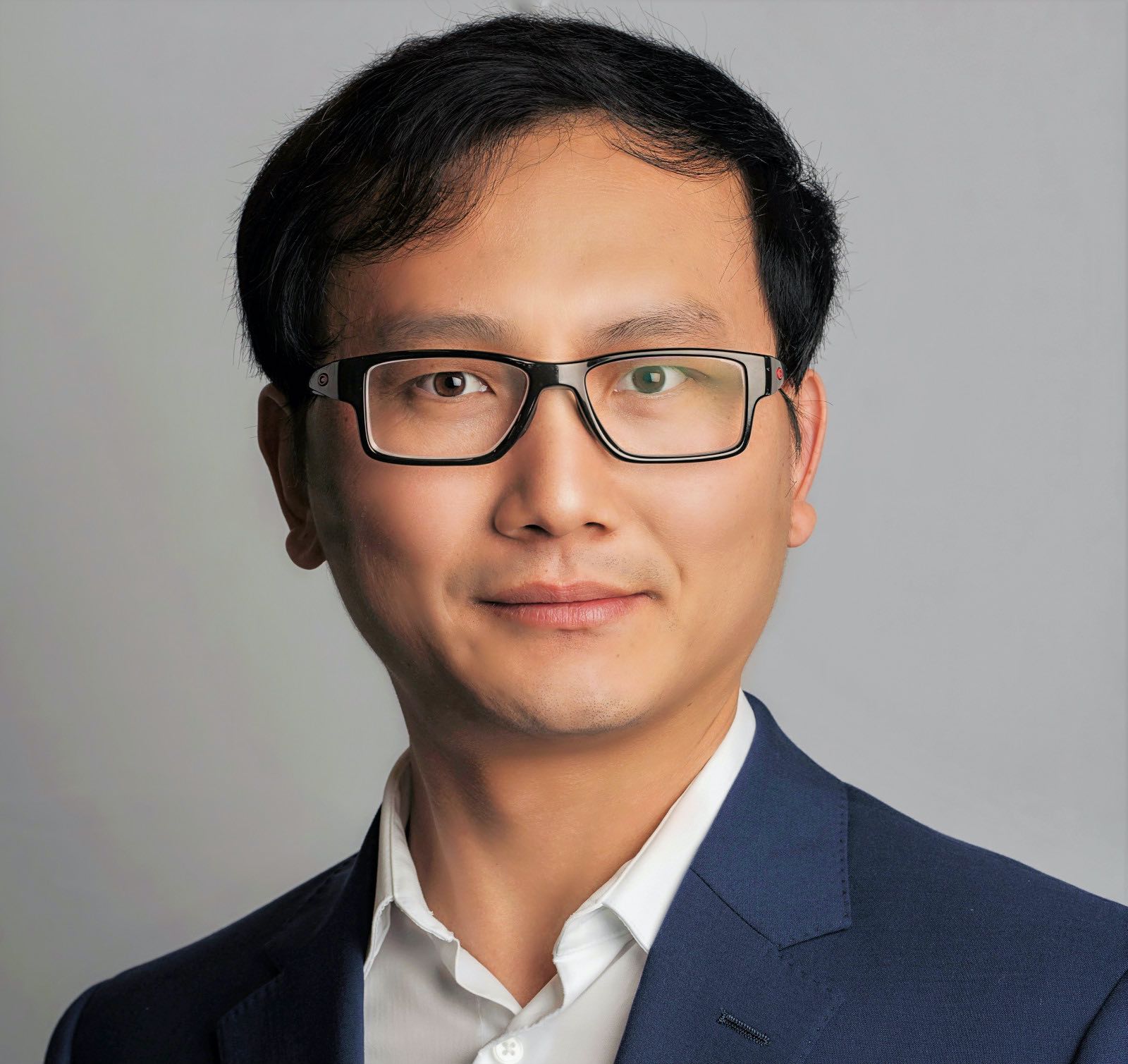}}]{Zhangyu Guan}
is an Assistant Professor with the Department of Electrical Engineering (EE) at The State University of New York at Buffalo (SUNY Buffalo). He received his Ph.D.\ in Communication and Information Systems from Shandong University in China in 2010. Dr. Guan was a visiting Ph.D.\ student with the Department of EE, SUNY Buffalo, from 2009 to 2010. He also worked at UB as a Postdoctoral Research Associate from 2012 to 2015. After that, he worked as an Associate Research Scientist with the Department of ECE at Northeastern University in Boston, MA, from 2015 to 2018. He directs the Wireless Intelligent Networking and Security (WINGS) Lab at SUNY Buffalo, with research interests in modeling, control, and system design toward next-generation, intelligent and secure wireless networking.
\end{IEEEbiography}

\vspace{-1cm}

\begin{IEEEbiography}
[{\includegraphics[width=1in,height=1.25in,keepaspectratio]{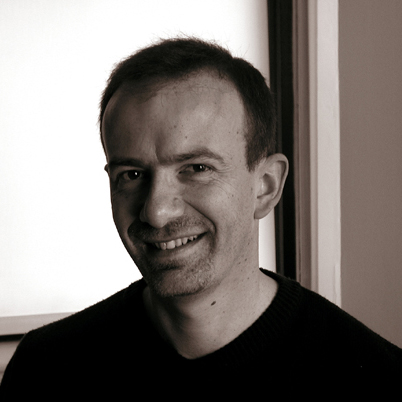}}]{Stefano Basagni}
is with the Institute for the Wireless Internet of Things and an associate professor at the ECE Department at Northeastern University, in Boston, MA. He holds a Ph.D.\ in electrical engineering from the University of Texas at Dallas (December 2001) and a Ph.D.\ in computer science from the University of Milano, Italy (May 1998). Dr.\ Basagni's current interests concern research and implementation aspects of mobile networks and wireless communications systems, wireless sensor networking for IoT (underwater and terrestrial), definition and performance evaluation of network protocols and theoretical and practical aspects of distributed algorithms. Dr. Basagni has published over nine dozen of highly cited, refereed technical papers and book chapters. His h-index is currently 44 (June 2020). He is also co-editor of three books. Dr. Basagni served as a guest editor of multiple international ACM/IEEE, Wiley and Elsevier journals. He has been the TPC co-chair of international conferences. He is a distinguished scientist of the ACM, a senior member of the IEEE, and a member of CUR (Council for Undergraduate Education).
\end{IEEEbiography}

\vspace{-1cm}

\begin{IEEEbiography}
[{\includegraphics[width=1in,height=1.25in,keepaspectratio]{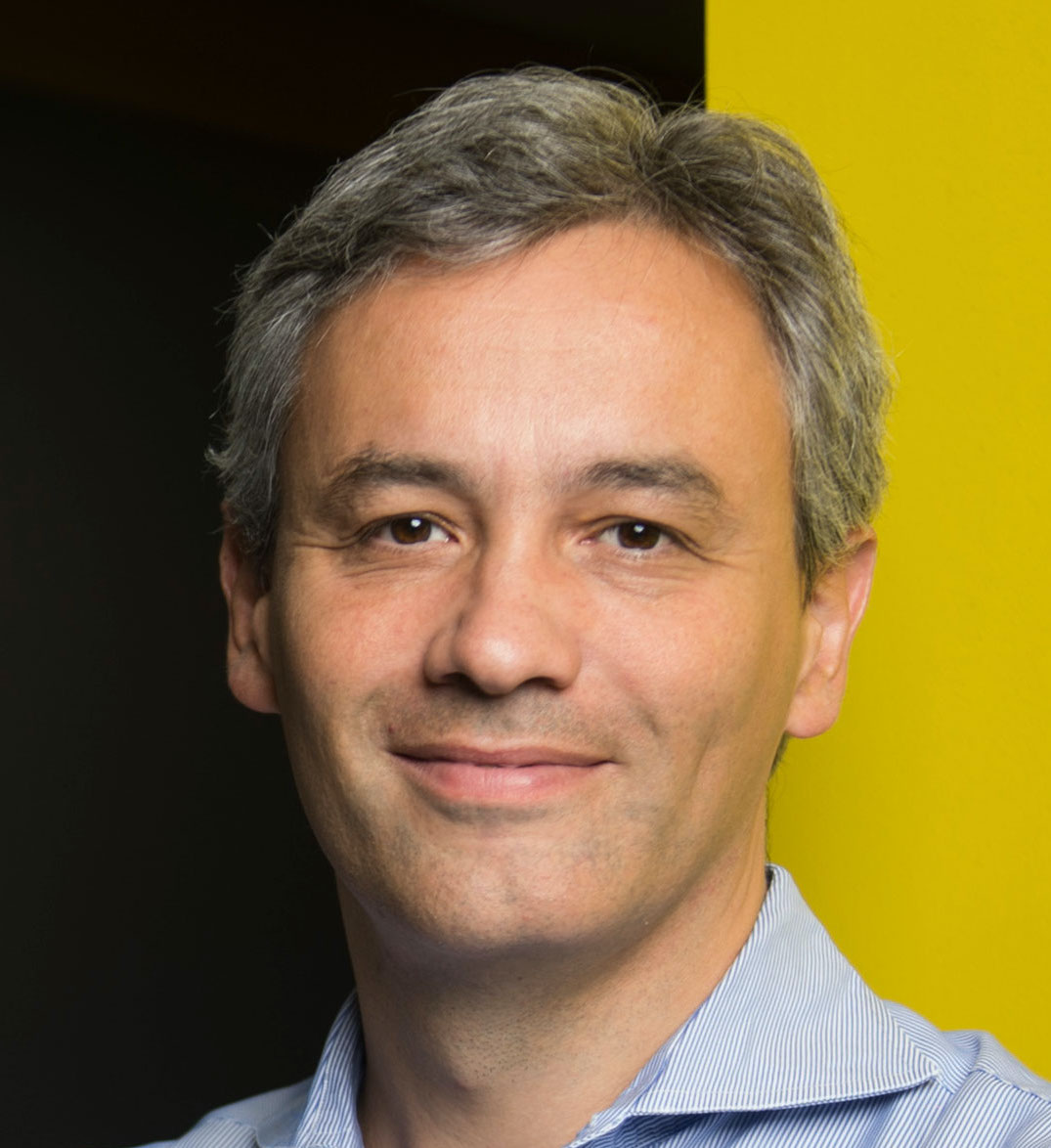}}]{Tommaso Melodia}
is the William Lincoln Smith Chair Professor with the Department of Electrical and Computer Engineering at Northeastern University in Boston. He is also the Founding Director of the Institute for the Wireless Internet of Things and the Director of Research for the PAWR Project Office. He received his Ph.D.\ in Electrical and Computer Engineering from the Georgia Institute of Technology in 2007. He is a recipient of the National Science Foundation CAREER award. Prof. Melodia has served as Associate Editor of IEEE Transactions on Wireless Communications, IEEE Transactions on Mobile Computing, Elsevier Computer Networks, among others. He has served as Technical Program Committee Chair for IEEE Infocom 2018, General Chair for IEEE SECON 2019, ACM Nanocom 2019, and ACM WUWnet 2014. Prof. Melodia is the Director of Research for the Platforms for Advanced Wireless Research (PAWR) Project Office, a \$100M public-private partnership to establish 4 city-scale platforms for wireless research to advance the US wireless ecosystem in years to come. Prof.\ Melodia's research on modeling, optimization, and experimental evaluation of Internet-of-Things and wireless networked systems has been funded by the National Science Foundation, the Air Force Research Laboratory the Office of Naval Research, DARPA, and the Army Research Laboratory. Prof. Melodia is a Fellow of the IEEE and a Senior Member of the ACM.
\end{IEEEbiography}

\end{document}